\documentclass[%
 reprint,
 amsmath,amssymb,
 aps,
]{revtex4-2}
\usepackage[utf8]{inputenc}
\usepackage{latexsym}
\usepackage{epsfig}
\usepackage{amsmath}
\usepackage{color}
\usepackage{hyperref}
\usepackage{mathrsfs}
\usepackage{graphicx}
\usepackage{mathtools}
\usepackage[amsmath]{empheq}
\usepackage{mathtools}
\usepackage{float} 
\pdfoutput=1

\newcommand{\be}{\begin{equation}}
\newcommand{\ee}{\end{equation}}
\usepackage{graphicx}
\usepackage{dcolumn}
\usepackage{bm}

\begin{document}

\preprint{APS/123-QED}

\title{Can the Universe decelerate in the future?}

\author{A. A. Escobal$^{1}$} \email{anderson.aescobal@gmail.com}
\author{J. F. Jesus$^{1,2}$}\email{jf.jesus@unesp.br}
\author{S. H. Pereira$^{1}$}\email{s.pereira@unesp.br}
\author{J. A. S. Lima$^{3}$}\email{jas.lima@iag.usp.br}

\affiliation{$^1$Departamento de F\'isica, Faculdade de Engenharia e Ci\^encias de Guaratinguet\'a, Universidade Estadual Paulista (UNESP),  Av. Dr. Ariberto Pereira da Cunha 333, 12516-410, Guaratinguet\'a, SP, Brazil. \\
\\$^2$Instituto de Ci\^encias e Engenharia, Universidade Estadual Paulista (UNESP),  R. Geraldo Alckmin, 519, 18409-010, Itapeva, SP, Brazil.\\
\\$^3$IAG, Universidade de S\~ao Paulo, 05508-900 S\~ao Paulo, SP, Brazil.
}

\author{}

\begin{abstract}
The possibility of an expanding decelerating  Universe in the distant future is investigated in the context  of a quintessence scalar field cosmology. Such a conceivable evolution is tested against SNe Ia and $H(z)$ {cosmic chronometers data}, and also through  a model independent method based on Gaussian Processes. The scalar field model is an extension of the exponential Ratra-Peebles (RP) quintessential cosmology whose potential now depends on a pair of parameters ($\alpha, \lambda)$ and predicts a decelerated expansion in the future. {  Different from RP approach the  $\alpha$ parameter allows for a decelerating  cosmology in the future thereby frustrating the inevitable evolution for a de Sitter Cosmology as predicted by the cosmic concordance model ($\Lambda$CDM).}  The statistical model analysis is updated with the most recent SNe Ia and $H(z)$ data thereby obtaining $H_0 = 68.6\pm3.7$ km/s/Mpc, $\Omega_{\Phi0} = 0.735^{+0.083}_{-0.069} $, $\alpha < 6.56$ and $\lambda< 0.879 $ (at $2\sigma$ c.l.). It is also found that the extended RP model allows for a future deceleration both for $H(z)$ and SNe Ia data. In the (model-independent) Gaussian Processes analysis, however, future deceleration is allowed only in the case of $H(z)$ data.
\vskip 0.2cm

\end{abstract}

\maketitle


\section{Introduction}
The current phase of accelerated expansion of the universe, initially confirmed in 1998 with observations of Type Ia Supernovae (SNe Ia) by two independent groups \cite{SN1,SN2} and reaffirmed by the most recent observations of the cosmic microwave background (CMB) radiation \cite{Planck2018}, is usually attributed to the domination of a material component with negative pressure thereby affecting the dynamics of the universe. This component  {corresponds to $\sim60-80\%$ of the whole material} content of the universe and does not interact with electromagnetic radiation, for this reason it is called Dark Energy (DE).

The flat $\Lambda$CDM cosmic concordance model assumes that DE is described by the cosmological constant $\Lambda$. This is the model that best describes most astronomical observations, both from the early universe, like nucleosynthesis \cite{BBN}, CMB \cite{Planck2018} and baryon acoustic oscillations (BAO) data \cite{SDSS:2005xqv}, and from the late time universe, such as measurements of SNe Ia and the Hubble parameter data, $H(z)$. However, it is well known that the flat $\Lambda$CDM model suffer from several observational limitations,  {among them the Supernova-CMB tension on the current values of the Hubble parameter ($H_0)$ \cite{ValentinoH0}, as well as the so-called $S_8$ tension \cite{ValentinoS8}. In this context, {if one excludes the possibility of unaccounted systematic effects \citep{NearbySupernovaFactory:2018qkd,Riess:2018kzi,Martinelli:2019krf},} alternative models are now required in order to solve both the theoretical and observational problems plaguing the $\Lambda$CDM model (see \citep{P2014,Riess:2020sih,Bull:2015stt} for detailed discussions about all the problems).}

Among several others, quintessence models arise as an alternative to describe the DE sector into the universe \cite{Martin:2008qp,Tsujikawa:2013fta}. This kind of model consider DE as a  {single minimally coupled real scalar field $\phi$} endowed with a certain associated potential $V(\phi)$. Scalar fields are widely used in several areas of physics, and particularly in cosmology they are applied to the whole evolution of the universe, since the grand unified theories (GUT) \cite{GUT,LCF}, inflation \cite{Sa:2020fvn,Lin:2009ta,Liddle:2008bm}, scalar field Dark Matter models \cite{Escobal,Magana012} and scalar field Dark Energy models \cite{Sahni:1999qe,Rubano:2001xi,Paliathanasis:2014zxa,Dimakis:2016mip}. This is due to its simple mathematical formalism, since all important information about the field is contained in the potential $V(\phi)$. 

In a quintessential cosmological model one can describe the different dynamic phases of the universe by determining the appropriate potential $V(\phi)$. Using the available observational data, it is also possible  to obtain  limits for the free parameters of the potential thereby constraining the specific model. An intuitive proposal to start studies with scalar fields is the simplest potential, namely the quadratic potential ($V(\phi) \propto \phi^2$), as done in \cite{Ratra:1990me,Urena-Lopez:2007zal}. A more general form of analysis for this kind of potential can be done by studying more general power laws potentials, like ($V(\phi) \propto \phi^n$), as discussed by Peebles and Ratra \cite{Peebles:1987ek,Ratra:1987rm}. In addition to these particular forms of potentials, many others have been adopted to describe quintessential cosmologies \cite{Yang:2018xah}.  Even other alternative candidates to accelerate the Universe, different from quintessence models, may also ultimately be described in terms of canonical scalar fields \cite{Basilakos:2013xpa,Lima:2012cm,Maia:2001zu}.

In this work we revisit the quintessence model proposed in \cite{Ademir}. This scalar field model is endowed with several interesting properties, among them: (i) at early times it behaves like a subdominant cosmological constant in a decelerating universe in perfect agreement with nucleosynthesis constraints, (ii) Such a decelerating phase is followed by an accelerating stage with quintessence dominance, and, finally, (iii) it allows a deceleration of the universe in the distant future. Beyond a scalar field approach, models with future deceleration can also be found from a more phenomenological viewpoint when the variable equation of state parameter of dark energy [$\omega(z)]$ goes to zero in the distant future (see, for instance, \cite{PDU2023} and Refs. therein). We also notice that a  possible  transition in the future for a decelerating stage thereby finishing naturally the eternal accelerating regime is a remarkable feature from a physical viewpoint. An eternal de Sitter phase as predicted by the $\Lambda$CDM model, for instance, is not in agreement with the requirements of S-matrix describing particle interactions \cite{Fischler:2001yj,Cline:2001nq}. 

In this context we seek to find stronger constraints to the free parameters of the quintessence model adopted here through the use of new observational data. By combining the SNe Ia data from the Pantheon sample \cite{Scolnic:2017caz} with the $H(z)$ data \cite{Magana:2017nfs} from cosmic chronometers,  we performed the statistical analysis of the model using Bayesian statistics thereby constraining  the free parameters of the proposed model (see also \cite{Guimaraes:2010mw} for an earlier model-independent analysis). Furthermore, we compared the results obtained by our analysis with the results from non-parametric methods, in this case the Gaussian Processes (GP) \cite{Seikel:2012uu,Holsclaw:2010nb,Shafieloo:2012ht, Jesus:2019nnk}, in order to obtain a better validation of the obtained results.

The paper is organized as follows. In Section \ref{dyn}, the theoretical foundations of the adopted scalar field cosmology plus CDM are briefly reviewed.  In Section \ref{data}, the observational data used in our analyses are described and, in Section \ref{result}, we present the  basic results derived here. We carry out the conclusions  and final remarks of our work in Section \ref{conc}. Finally, some technicalities including  the most relevant accounts describing the model can be seen in the Appendix.

\section{\label{dyn} Cosmological Model}

Let us now determine the set of equations driving the dynamics of the quintessential cosmological model. The dark energy component is described by a homogeneous canonical scalar field $\phi$ with energy density and pressure:
\begin{align}\label{eqrho}
     \rho_{\phi}&= \dfrac{1}{2}\dot{\phi}^2 + V(\phi)\,,\\\label{eqp}
      p_{\phi}& = \dfrac{1}{2}\dot{\phi}^2 - V(\phi)\,.
\end{align}
The potential $V(\phi)$ contains all the physical information about the field. The independent Friedmann equations take the following form:
\begin{align}
    H^2 &= \dfrac{8 \pi G}{3}\left(\rho_m + \rho_{\phi}\right) -\dfrac{k}{a^2}\,,\\
     \dfrac{\ddot{a}}{a} &= - \dfrac{4\pi G}{3}\left(\rho_m+\rho_\phi+3p_\phi \right)\,,
\end{align}
where $H = {\dot a}/a$ is the Hubble parameter.  {From now on, we shall work in an spatially flat Universe ($k=0$), as indicated by Planck \cite{Planck2018} and required by inflation \cite{MartinEtAl13}.} The equation of motion obeyed by the scalar field which is also contained in the energy conservation law ($u_{\mu}T_{(\phi);\nu}^{\mu\nu} = 0$) can be written as: 
\be\label{EM1}
    \ddot{\phi} + 3H\dot{\phi} + \dfrac{d\,V(\phi)}{d \phi} = 0\,.
\ee

{Now, in order to simplify our calculations, let us introduce the dimensionless field $\Phi$ and the dimensionless potential $U(\Phi)$ defined in terms of $\phi$ by the following expressions: 
\begin{align}\label{PHiDef}
    \Phi &= \sqrt{\dfrac{8\pi G}{3}}\phi\,,\\ \label{UDEF}
    U(\Phi) &= \dfrac{8\pi G}{3H_0^2}V(\Phi)\,,
\end{align}
which are explained with more detail} in the Appendix \ref{ap2}.

The potential $U(\Phi)$ adopted in our analysis follows  directly from the work \cite{Ademir}  (see discussion below their equation (4) and also \eqref{B28} in the Appendix).  It is given by
\begin{align}
   U(\Phi)    &= \Omega_{{\Phi0}}e^{-\frac{1}{2}\left[3\alpha \Phi^2+2\sqrt{3\lambda}\Phi\right]}\left[1-\dfrac{\lambda}{6}\left( \alpha\sqrt{\dfrac{3}{\lambda}}\Phi +1\right)^2 \right]\,,
\end{align}
where $\alpha$ and $\lambda$ are constants proposed in the approximation made in \cite{Ademir} (see their  equation \eqref{APADM}). As should be expected, the above expression in the limit $\alpha \rightarrow 0$ can be seen as the dimensionless potential associated to the Ratra-Peebles model \cite{Ratra:1987rm}. 

{At this point, it is also convenient to change the time coordinate $t$ to the redshift $z$ through the standard transformation:
\begin{align}
    \dfrac{d}{d t} = -H(1+z)\dfrac{d}{dz}\,.
\end{align}
In this way, the equation of motion (\ref{EM1}) becomes:
\begin{widetext}
\begin{align}\label{eqmvC}
       \Phi'' + \left[ 3\Omega_{m0}(1+z)^2+\left(3(1+z)\Phi'^2-\dfrac{4}{1+z} \right)\right]\dfrac{\Phi'}{2E^2}+ \dfrac{1}{E^2(1+z)^2}\frac{d\,U(\Phi)}{d \Phi}  &= 0\,,
\end{align}
\end{widetext}
where primes denote derivatives $d/dz$ while the dimensionless quantity $E$ is defined in the usual manner, $E^2 \equiv {H^2}/{H_0^2}$, which in the present context takes the form:
\begin{align}\label{E2U}
      E^2 &= \Omega_{m0}(1+z)^3+ \frac{2U(\Phi)}{1-w}\,,
\end{align}
where $\Omega_{m0}$  is  the present day matter density parameter (hereafter a subindex  ``0" will denote a present time quantity).} Being $w$ the parameter of the equation of state (EoS) of the form
\begin{align}
    p_{\phi} &= w \rho_{\phi}\,,\label{eqwww}
\end{align}
it is easy to obtain the expression for the parameter $w$:
\begin{align}
         w &= \dfrac{\Omega_{m0}(1+z)^5\Phi'^2+2U(\Phi)\left[(1+z)^2\Phi'^2-1 \right] }{\Omega_{m0}(1+z)^5\Phi'^2+2U(\Phi)}  \,.
\end{align}

In the same vein, it is also possible to rewrite the deceleration parameter $q(t)$ 
\begin{align}
    q = -\dfrac{\ddot{a}}{aH^2}\,,
\end{align}
in terms of the redshift $z$, this may be written as:
\begin{align}
     q(z)=  \frac{3\left[\Omega_{m0}(1+z)^5\Phi'^2+2U(\Phi)\left((1+z)^2\Phi'^2-1 \right) \right]}{4\Omega_{m0}(1+z)^3+4U(\Phi)}+\dfrac{1}{2}\,.\label{qzdef}
\end{align}

Let us now determine an equation that allows us to analyse the SNe Ia data,  { which depends on the dimensionless luminosity distance $D_L$ given in terms of the dimensionless comoving distance, $D_C$ as}
\begin{equation}
D_L=(1+z)D_{C}\,.
\end{equation}
 {where dimensionless distances $D_i$ relate to dimensionful distances $d_i$ as: \begin{align}
D_i\equiv\frac{d_i}{d_H},
\end{align}
where $d_H\equiv\frac{c}{H_0}$ is Hubble distance.}
 Now, since $ D_C$ depends on $ E(z) $ but we do not have an analytic expression for $ E (z) $, we need a differential equation for $ D_C$. For a spatially flat universe we can write:
\begin{equation}
\label{clark}
\frac{dD_{C}}{dz}\equiv \frac{1}{E(z)}\,.
\end{equation}

Thus, with the equation of motion of the field $\Phi$ given by \eqref{eqmvC} and with the equation \eqref{E2U} that we have just determined, we have all the equations necessary to start the statistical analysis. Let us then determine the initial conditions of the field $\Phi $ and its derivative $\Phi'$.  {As shown in the Appendix \ref{ap2}, we consider the initial condition for the field currently to be}
\begin{align}\label{Phi0}
    \Phi_0= \Phi(z = 0) = 0\,.
\end{align}
From the value of $\Phi_0$, we can determine $\Phi'$ through the equation \eqref{EPHI}, since $E^2(z=0) = 1$, we then have
\begin{align}\label{Phi'0}
         \Phi'_0 &=-\sqrt{ \dfrac{\lambda}{3}\Omega_{\Phi0}}\,.
\end{align}

Now we have all the tools to perform the numerical analysis for the model using the $H(z)$ and SNe Ia observational data.

\section{\label{data}Cosmological Data}

As remarked in the introduction, the observational dataset used in this work consists of two independent classes of astronomical observations, namely: (i) the compilation of SNe Ia from the Pantheon sample \cite{Scolnic:2017caz}, and (ii) the latest measurements of the Hubble parameter, $H(z)$ \cite{Magana:2017nfs}.

The Pantheon sample has $1048$ data from SNe Ia, within the redshift range $0.01\,<\, z\,<\,2.3$, containing measurements of SDSS, Pan-STARRS1 (PS1), SNLS, and various HST and low-$z$ datasets.

\ 

{The idea behind SNe Ia is that they work as standard candles, in the sense that they have nearly the same luminosity when the supernova event occurs. This is due to the fact that the explosion occurs when the dwarf star in the binary system reaches always the same mass, namely, the Chandrasekhar mass limit. However, due to differences in environment, color etc., they are not exactly standard candles but they are really standardizable. The process of standardizing SNe Ia involves calibrations that are independent of cosmological models, based just on astrophysical assumptions, as the reddening due to dust, for instance \cite{Tripp98,Scolnic:2017caz}.}

{The Hubble $H(z)$ parameter data used in our analysis covers a redshift range of $0.07 < z < 1.965$. We have used the most complete sample of $H(z)$ measurements, with 31 data, obtained by estimating the differential age of galaxies \cite{Simon2004, Stern2009, Moresco2012, Zhang2012, Moresco2015, Moresco2016}, usually dubbed cosmic chronometers. It is interesting to use this observational dataset here because they are obtained through astrophysical assumptions only, as the luminosity of the main-sequence turnoff \cite{JimenezEtAl03}, being independent of the choice of the background cosmological model. The idea basically is to obtain ages of extragalactic globular clusters, then use these ages to obtain an envelope in the age-redshift relation. This can be used to estimate $\frac{dz}{dt}$ and finally to estimate $H(z)=-\frac{1}{1+z}\frac{dz}{dt}$, without the necessity of a specific expanding  model.}

{Our analysis consists of two steps: First of all, we make a parameter estimation of the scalar field model. Secondly, we compare this analysis with a model independent reconstruction, which is the GP method. In the first step, we choose to work only with SNe Ia and $H(z)$ because they are independent of cosmological model assumptions and so we can make a comparison with the model-independent GP method.}

\begin{figure*}
    \centering
    \includegraphics[width=0.49
    \textwidth]{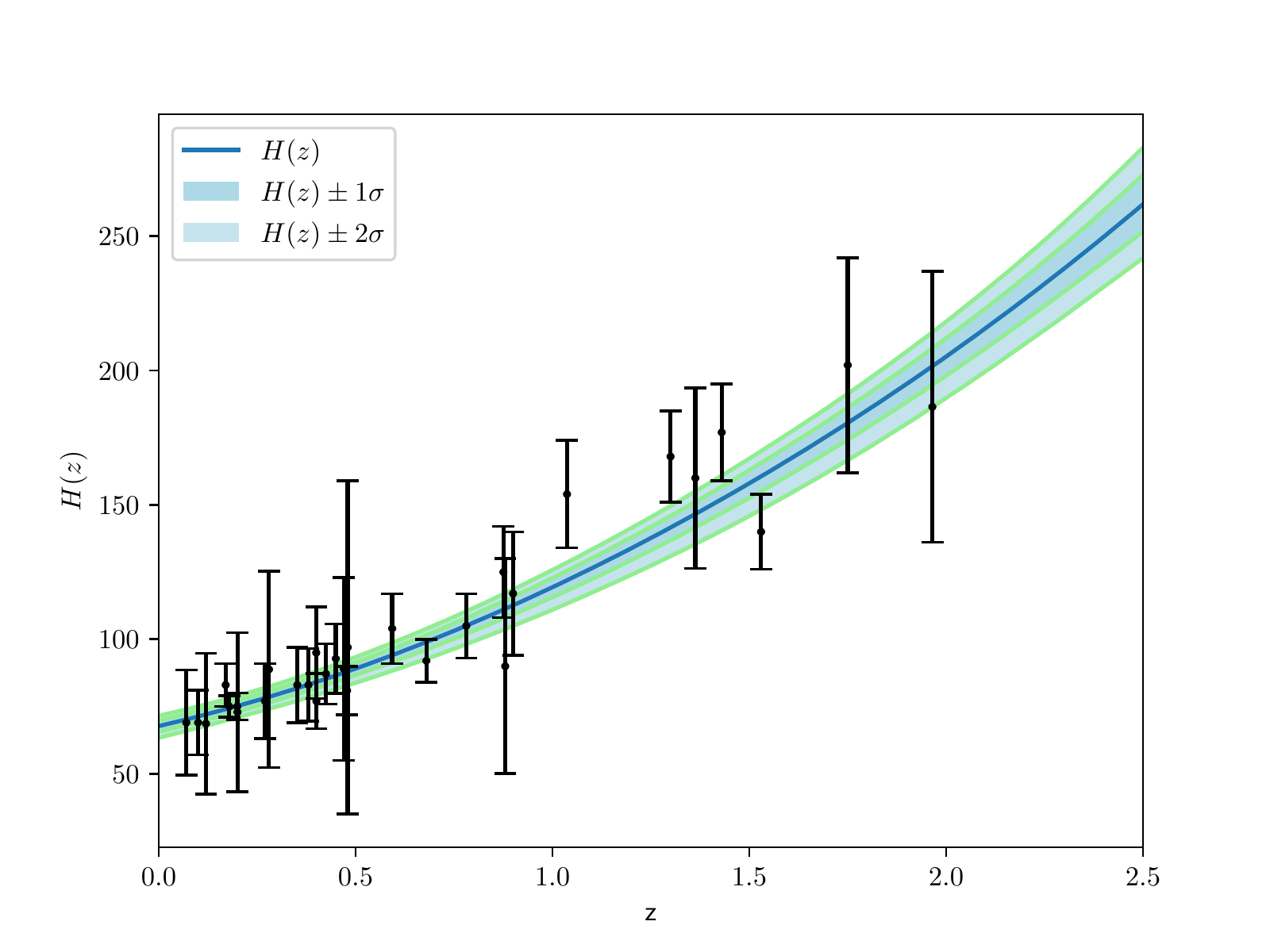}
    \includegraphics[width=0.49
     \textwidth]{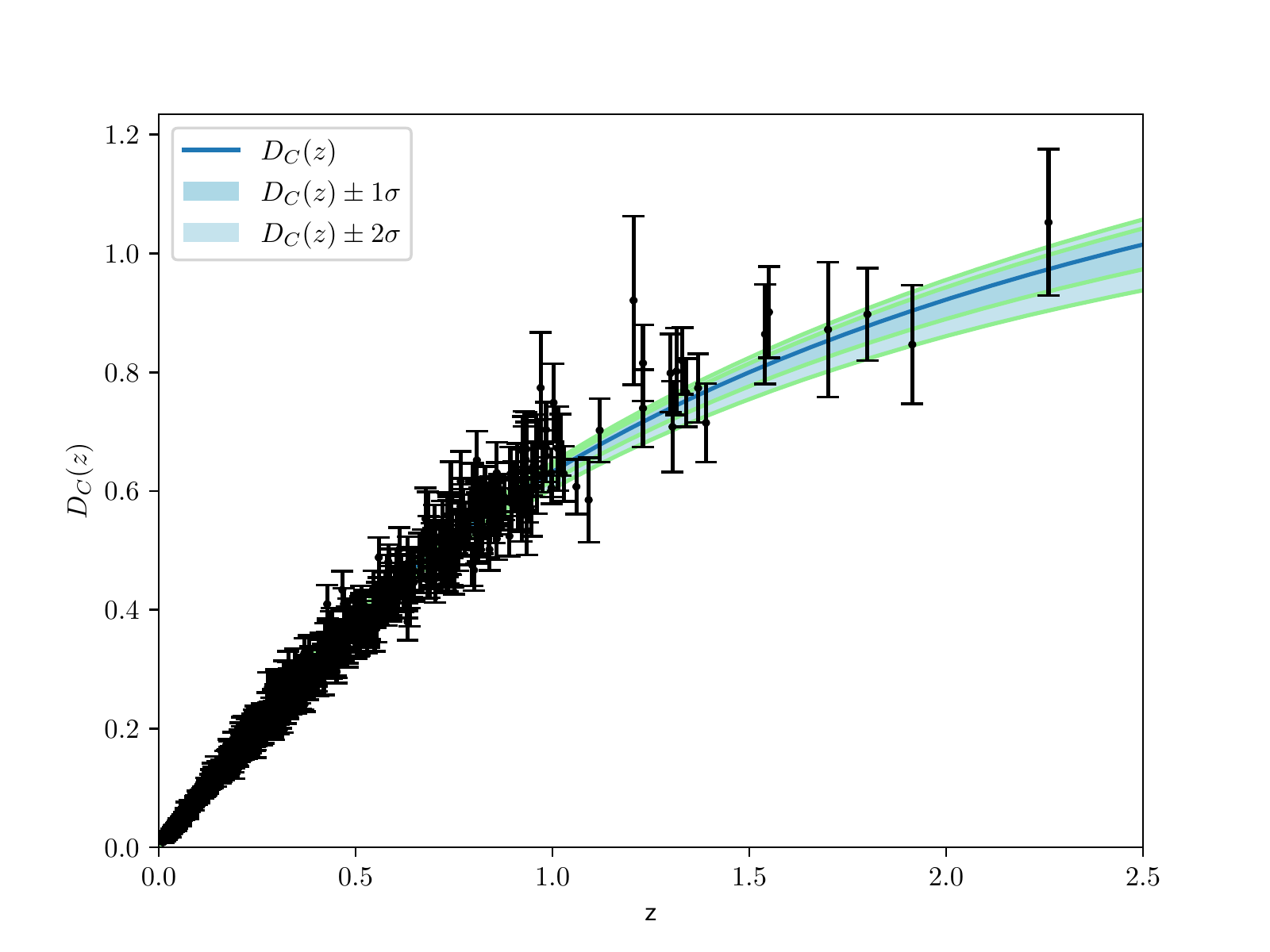}
       \caption{\textbf{Left:} Plot of the 31 CC data along with the $H(z)$ curve from the best fit model and 1 and 2$\sigma$ confidence intervals. $H(z)$ is in units of km/s/Mpc. \textbf{Right:} Plot of $D_C(z)$ taken from the Pantheon sample together with the $D_C(z)$ curve from the best fit model and 1 and 2$\sigma$ confidence intervals. {The best fit model and the confidence intervals shown correspond to the mean values of the parameters and errors of Tab. \ref{tab1}.}}
            \label{dados}
\end{figure*}

In Fig. \eqref{dados}, we plot the observational data used in this work. The $31$ CC data are shown in the left panel, jointly with the $H(z)$ function of the analyzed model at a confidence interval of $2\sigma$. The right panel, on the other hand, shows the dimensionless comoving distances obtained with the Pantheon sample jointly with the $D_C(z)$ 2$\sigma$ confidence intervals of the model. {The best fit model and the confidence intervals shown correspond to the mean values of the parameters and errors of Tab. \ref{tab1}.}

\section{\label{result}Analyses and Results}

We have used Bayesian statistics in order to find the values of the free parameters of the model. Using a flat prior $\pi$ over the parameters with a likelihood of the form $\mathcal{L}\propto e^{-\chi^2/2}$, we can write the posterior probability distribution $p\propto \pi \mathcal{L}$.  {The prior used on the parameters is shown in the table \eqref{tab:priors}}, we have used the following flat priors: $H_0\in[50,100]$, $\Omega_{\phi0}\in[0,1]$, $\alpha\in[0,10]$, $\lambda\in[0,5]$
\begin{table}[ht]
    \centering
    \begin{tabular} {| c|  c|}
    \hline
 Parameter &  Flat prior interval\\
\hline
{\boldmath$H_0$} (km/s/Mpc) & $[20,120]$\\
{\boldmath$\Omega_{\phi0}       $} & $[0,1]$\\
{\boldmath$\alpha         $} & $[0,10]$\\
{\boldmath$\lambda       $} & $[0,5]$\\
\hline
\end{tabular}
    \caption{Chosen priors for the free parameters.}
    \label{tab:priors}
\end{table}

It is worth to mention that we have chosen to work in the interval $\alpha>0$ due to divergences in the potential in the region $\alpha<0$. { In addition, the interval $\lambda>0$ has also been selected in order to have a real-valued $\frac{d\phi}{da}$ as can be seen in Eqs. \eqref{dphida} and \eqref{APADM}.}

By sampling the probability functions of the combination of the data of $H(z)$ and SNe Ia, {using a simple and powerful MCMC method called Affine Invariant MCMC Ensemble Sampler \cite{GoodWeare}, which was implemented in the {\sffamily Python} language with the {\sffamily emcee} \cite{ForemanMackey13} software.}  {The convergence of the chains is obtained using the autocorrelation time ($\tau$) provided by the {\sffamily emcee} software. As explained in the {\sffamily emcee } documentation \footnote{\url{https://emcee.readthedocs.io/en/stable/}}, a good estimate of $\tau$ is obtained when $n_{\mathrm{samples}}\gg\tau$ (where $n_{\mathrm{samples}}$ is the number of MCMC samples from each walker and we have used 100 walkers). We found $n_\mathrm{samples}>50\tau$ for all free model parameters. As suggested in the {\sffamily emcee } documentation, we have discarded as burn-in $\sim2\tau$ samples and have thinned the chains at each $\sim\tau/2$ samples.}
 
 Constraints are plotted on the same figure, Fig. \ref{Tplot}. We use the freely available software {\sffamily getdist} \cite{Lewis19}
 , in its {\sffamily Python} version. We were able to determine the region with the highest probability of finding the values of the free parameters of the model, as shown in the Figure \ref{Tplot}, with contours  {corresponding to $1\sigma$ and $2\sigma$ (68\% and 95\% c.l.).

\begin{figure*}
    \centering
    \includegraphics[width=0.9\textwidth]{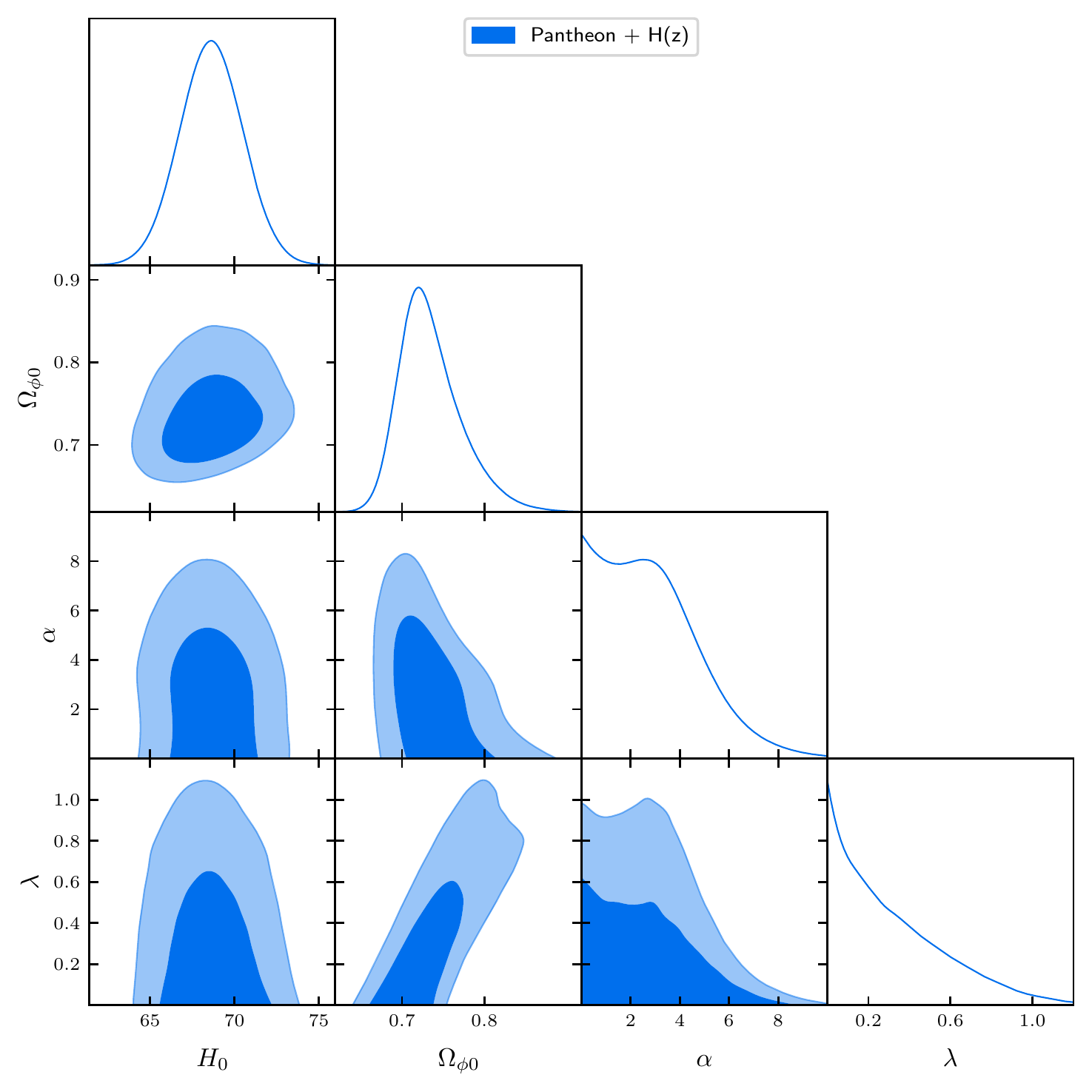}
    \caption{Triangular plot of parameters, with data from $H(z)$ and SNe Ia combined.  {The contours correspond to 68\% and 95\% c.l.}}\label{Tplot}
\end{figure*}

We show in the Table \ref{tab1} the values of the free parameters of the model in a confidence interval of $95\%$. We get $\Omega_{\Phi0} = 0.735^{+0.083}_{-0.069}$. The constraints obtained for the pair of parameters ($\alpha, \lambda$), defining  the scalar field potential are $\alpha < 6.56 $ and $\lambda < 0.879$, while the Hubble constant for the model is $68.6\pm3.7$ km/s/Mpc. {This $H_0$ value is similar to what is obtained in the same analysis, in the context of flat $\Lambda$CDM model \cite{ParkRatra18}, $H_0=69.1\pm1.8$ km/s/Mpc ($1\sigma$ c.l.)}
 
\begin{table}[H]
    \centering
\begin{tabular} { l  c}
\hline
 Parameter &  95\% limits\\
\hline
{\boldmath$H_0$ \textbf{(km/s/Mpc) }         } & $68.6\pm3.7        $\\

{\boldmath$\Omega_{\phi0}$} & $0.735^{+0.083}_{-0.069}   $\\

{\boldmath$\alpha$       } & $< 6.56                   $\\

{\boldmath$\lambda$      } & $< 0.879                  $\\
\hline
\end{tabular}
\caption{Mean values of the free parameters.}
\label{tab1}
\end{table}

With the sample of parameters generated by emcee, we  {have plotted} the deceleration parameter $q(z)$ in a confidence interval of $2\sigma$ as shown in the left panel of Figure \ref{Uplot}, where we expanded $ q(z )$ to $z \approx -0.5$. Thus, we can see that the model allows $q(z)>0$ within $1\sigma$ confidence for $z\lesssim-0.4$, approximately. With the same data sample in the right panel of Figure \ref{Uplot}, we plot the dimensionless potential $U(z)$, which is also displayed at a confidence interval of $2\sigma$.  {Here, we should mention that the confidence intervals that are shown in Fig. \ref{Uplot} were obtained as if $q(z)$ and $U(z)$ were derived parameters, with the following method: for a fixed redshift, let us say, $z=0$, one obtains the chain for $q(z)$ (or $U(z)$) from the chains of the free (primitive) parameters, indicated in Tab. \ref{tab1}. In this way, any correlations between the parameters, as well as any asymmetries coming from their distributions will be taken into account in the determination of $q(z)$ and $U(z)$.}
\begin{figure*}
    \centering
    \includegraphics[width=0.49
    \textwidth]{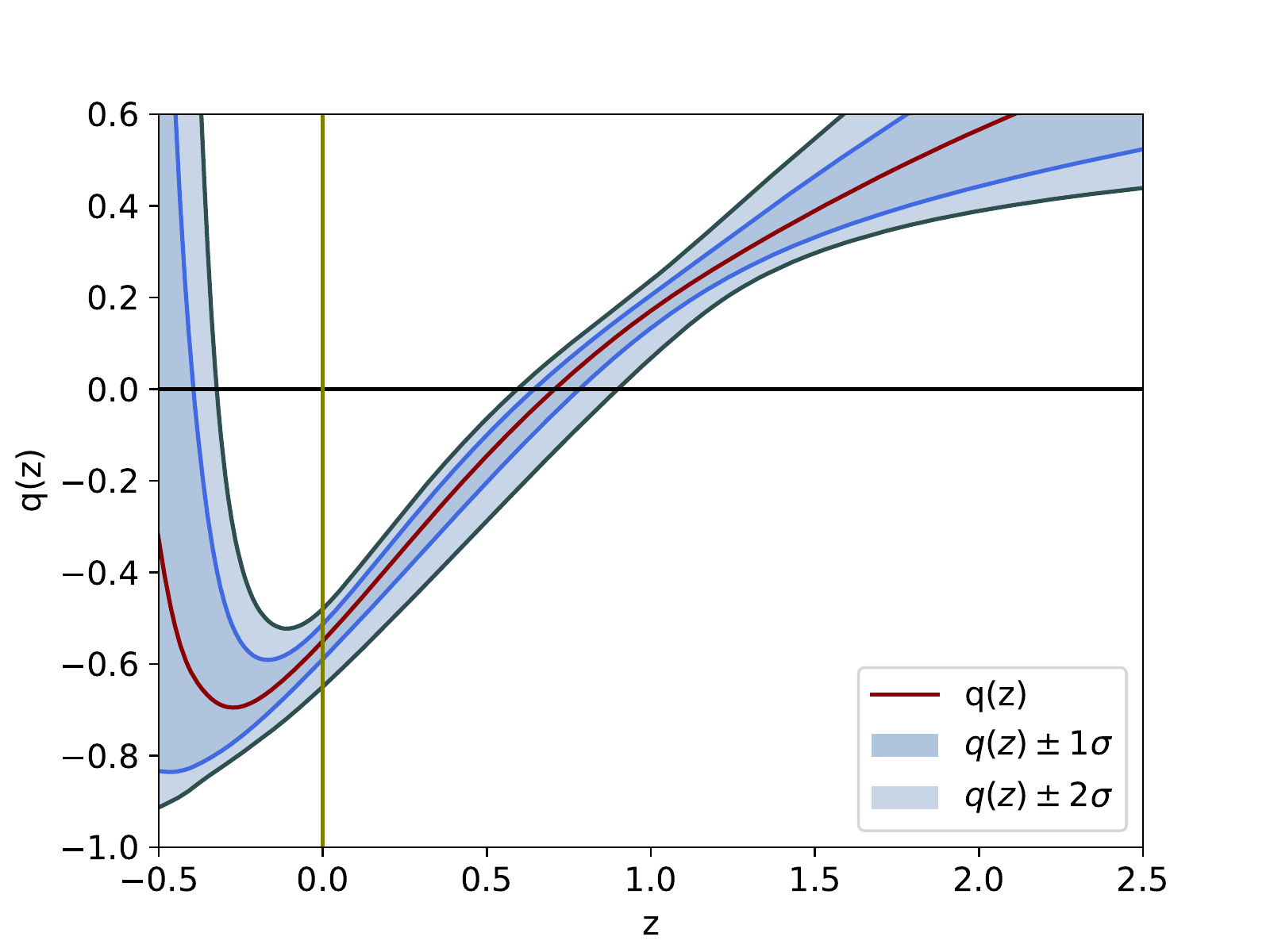}
    \includegraphics[width=0.49
     \textwidth]{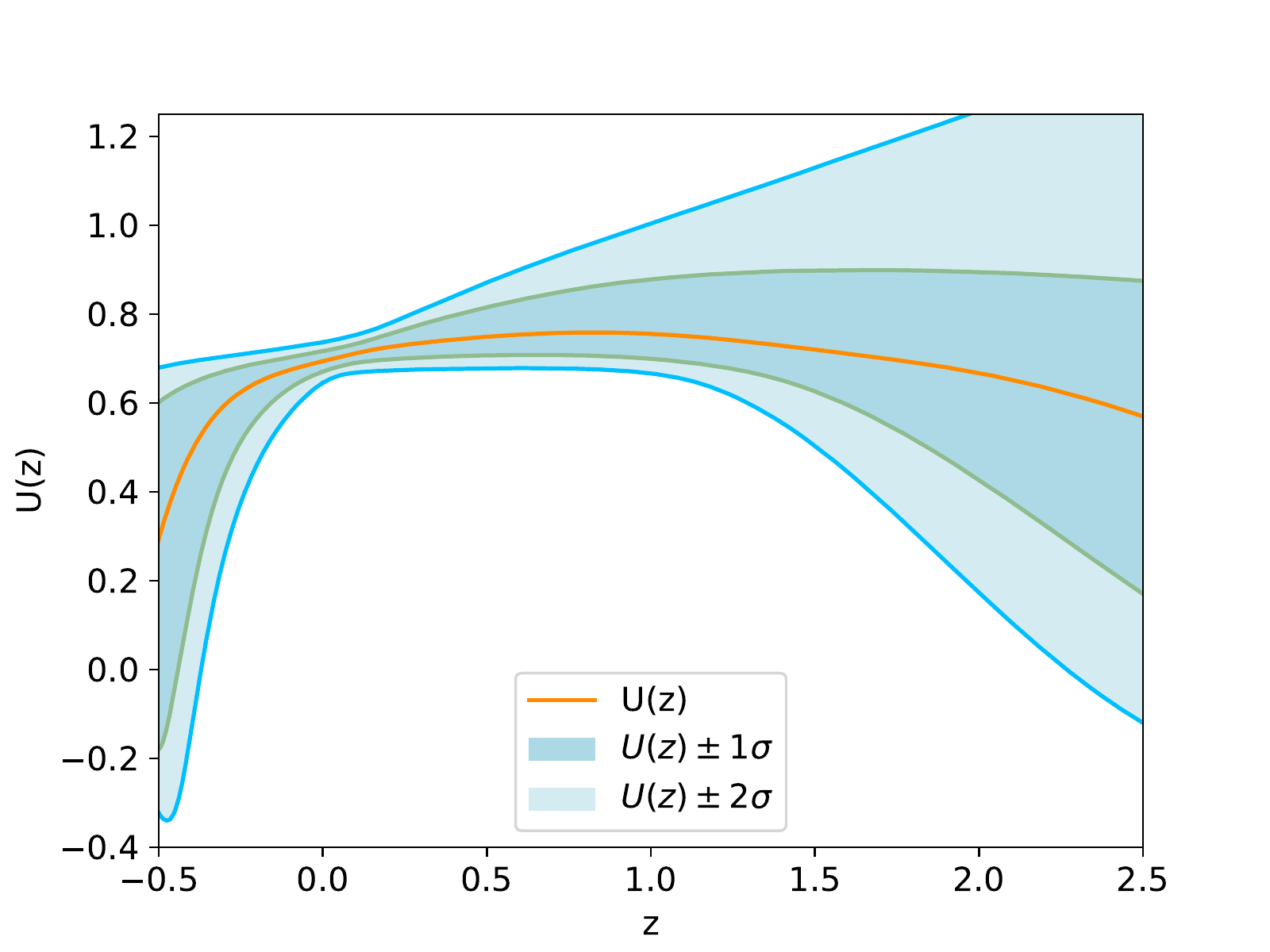}
       \caption{\textbf{Left:} Plot of $q(z)$ using the  {chains of the free parameters generated by emcee. Both reconstructions correspond to the joint analysis of Pantheon+$H(z)$ data.} \textbf{Right:} Plot of $U(z)$ using the {chains of the free parameters generated by emcee.} }
            \label{Uplot}
\end{figure*}

From the non-parametric method GP \footnote{See \cite{Seikel:2012uu} and \cite{Jesus:2019nnk} for more details on Gaussian processes}, using the SNe Ia and $H(z)$ data, the reconstruction of the dimensionless potential $U(\Phi)$ was done in \cite{Urec}, for a generic quintessence model. In \cite{Jesus:2019nnk}, the deceleration parameter $q(z)$ was reconstructed. Based on such analyses, we will expand the reconstruction of $q(z)$ and $U(z)$ to the future time, that is, in the redshift range of $-1 < z < 0$, in order to find a non-parametric result to compare with the analyzed model.  {The reconstructions were obtained using a correlation function (kernel)\cite{Jesus:2019nnk} Exponential Square, $k(x_i,x_j)$, between points $x_i$ and $x_j$ of the data sample. The Exponential Square kernel is given by:
   \begin{align} k(x_i,x_j)=\sigma_f^2\exp\left[-\frac{(x_i-x_j)^2}{2\theta^2}\right]\,,
   \end{align}
where $\theta$ and $\sigma_f$ are the GP hyperparameters obtained from the data.} The $q(z)$ reconstruction can be obtained from the observables $H(z)$ and $D_C(z)$ as:
\begin{align}\label{qzGP}
    q(z)=(1+z)\frac{H'}{H}-1=-(1+z)\frac{D_C''}{D_C'}-1\,.
\end{align}

As explained in Ref. \cite{Urec}, $U(z)$ can be obtained, assuming spatial flatness, from the observables as:
\be
U(z)=E^2-\frac{E(1+z)}{3}\frac{dE}{dz}-\frac{\Omega_{m0}(1+z)^3}{2}\,.
\label{UphiEz}
\ee
\be
U(z)=\frac{1}{D_C'^2}+\left(\frac{1+z}{3}\right)\frac{D_C''}{D_C'^3}-\frac{\Omega_{m0}(1+z)^3}{2}\,. \label{UzDc}
\ee  

{As one can see from these equations, $U(z)$ depends on the parameter $\Omega_{m0}$, which can not be obtained from the GP method alone. As our idea with the GP method is to obtain reconstructions which are the most model independent as possible, we choose to work with a large prior over $\Omega_{m0}$, namely, $\Omega_{m0}=0.30\pm0.05$, which was also used in \cite{Urec}.}

We have implemented this Gaussian Process method, in order to obtain model independent reconstructions of $q(z)$ and $U(z)$ from the data alone, by using the freely available package GaPP \cite{Seikel:2012uu} \footnote{GaPP is currently available in the auxiliary repository \url{https://github.com/carlosandrepaes/GaPP}}. The hyperparameters $\theta$ and $\sigma_f$ were obtained by optimization of the GP marginal likelihood. {The values obtained for the hyperparameters $(\sigma_f,\theta)$ were, respectively, $(133.75, 1.93)$ in the case of $H(z)$ reconstruction and $(9.11\times10^4, 2.51)$ in the case of $D_C(z)$ reconstruction from SNe Ia data.}

\begin{figure*}
    \centering
    \includegraphics[width=0.49\textwidth]{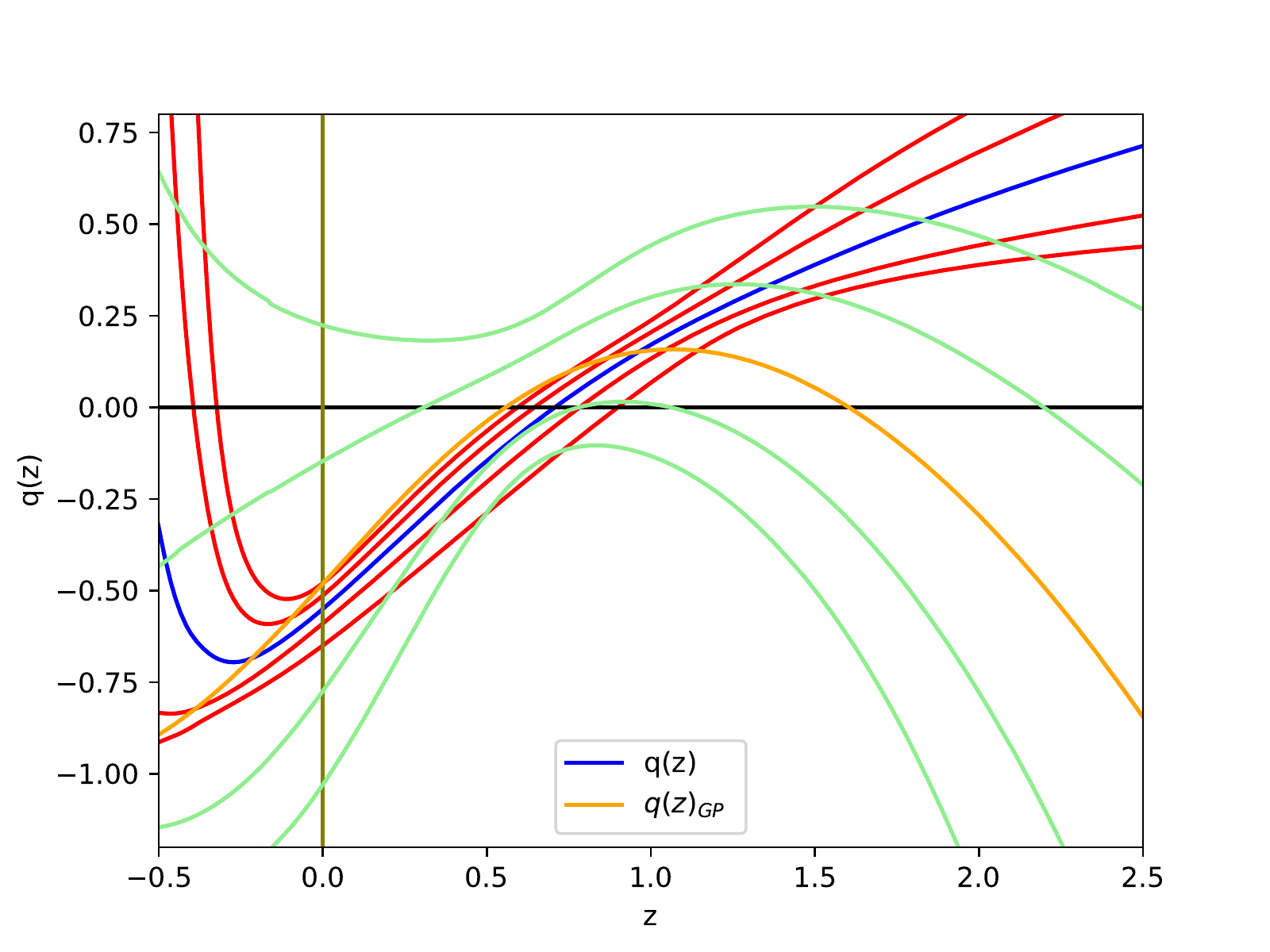}
     \includegraphics[width=0.49\textwidth]{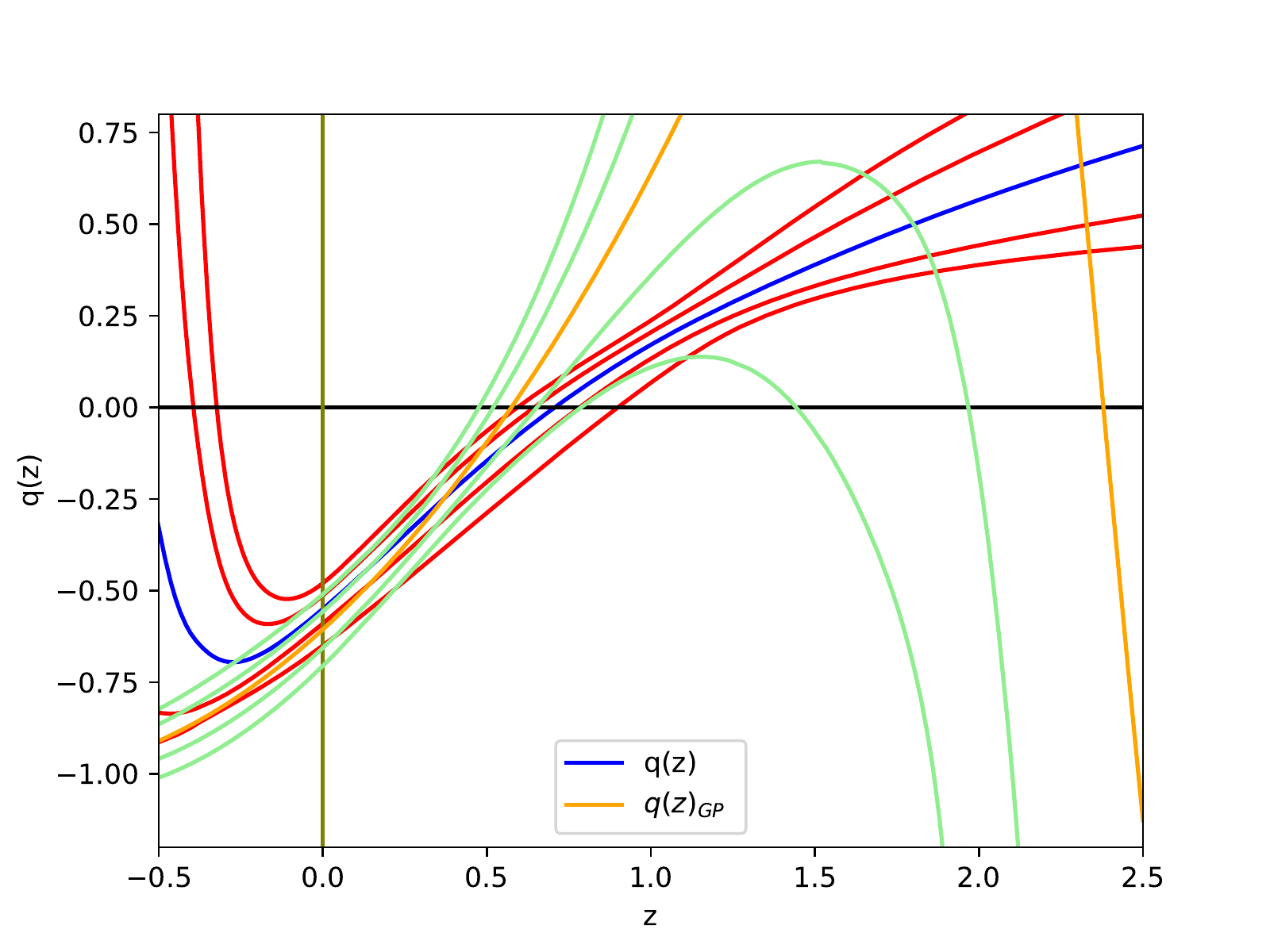}
       \caption{Plot of the $q(z)$ of the analysed model together with $q(z)$ reconstructed by GP.  {Also shown are the $1\sigma$ and $2\sigma$ confidence intervals, in red for the model and in green for the GP reconstructions.} \textbf{Left:} GP reconstruction from $H(z)$ data. \textbf{Right:} GP reconstruction from Pantheon.}
       \label{qRec}
\end{figure*}

In Figure \ref{qRec}, we plot the reconstruction of $q(z)$ via GP and the numerical result of $q(z)$ as predicted by the investigated model.  On the left we show the reconstruction of $q(z)$ via GP obtained from the $H(z)$ data. We see that the reconstruction of $q(z)$ reaches positive values for $z<0$ being less than $1\sigma$ compatible with the $q(z)$ of the analyzed model. On the right we present the reconstruction of $q(z)$ via GP using the SNe Ia data and we also show the $q(z)$ for the studied model. Although the $q(z)$ found to match $1\sigma$ over most of the redshift range we do not get values of $q(z) > 0$ for the reconstruction from the SNe Ia data.

\begin{figure*}
    \centering
    \includegraphics[width=0.49
    \textwidth]{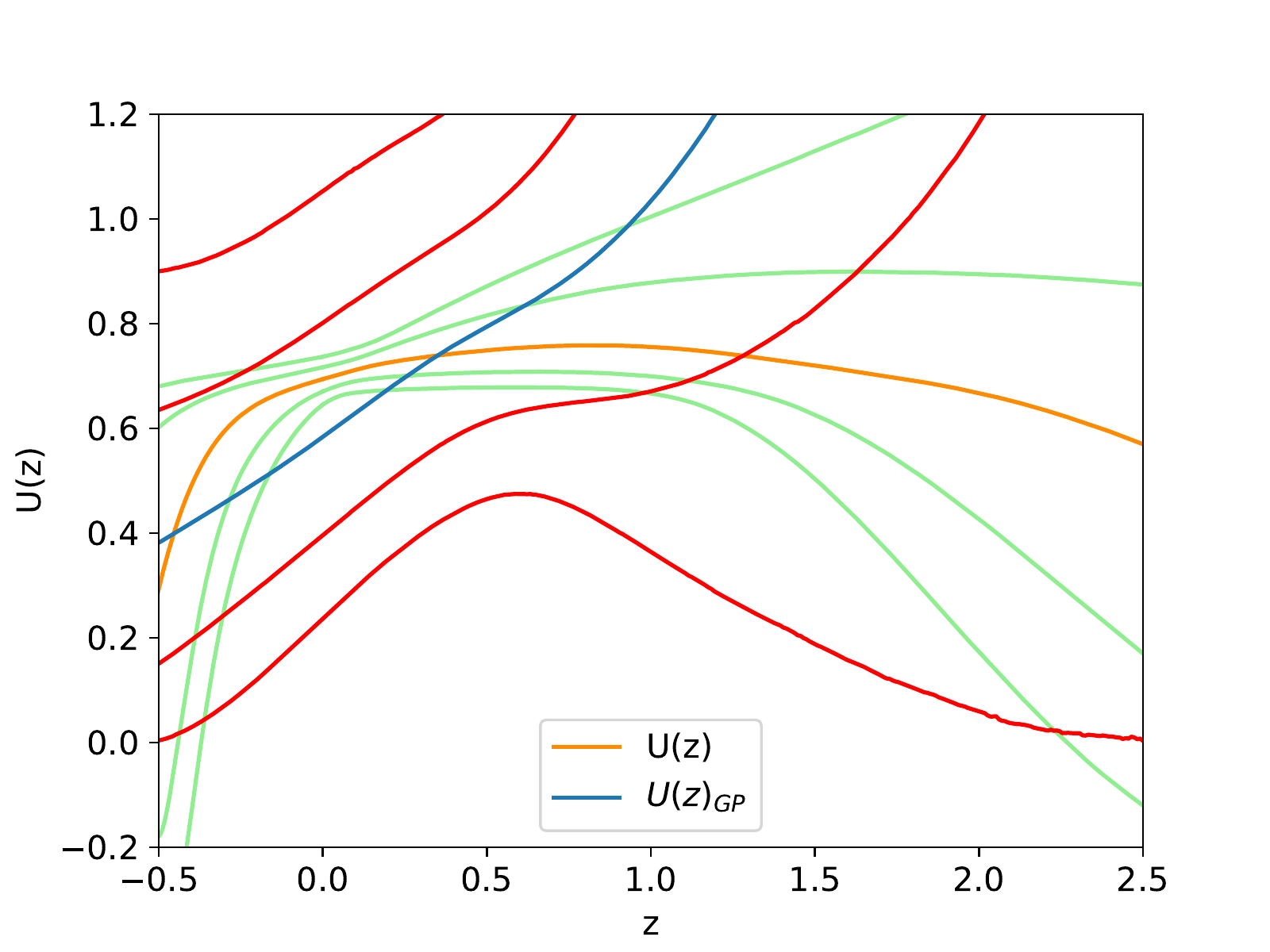}
     \includegraphics[width=0.49
     \textwidth]{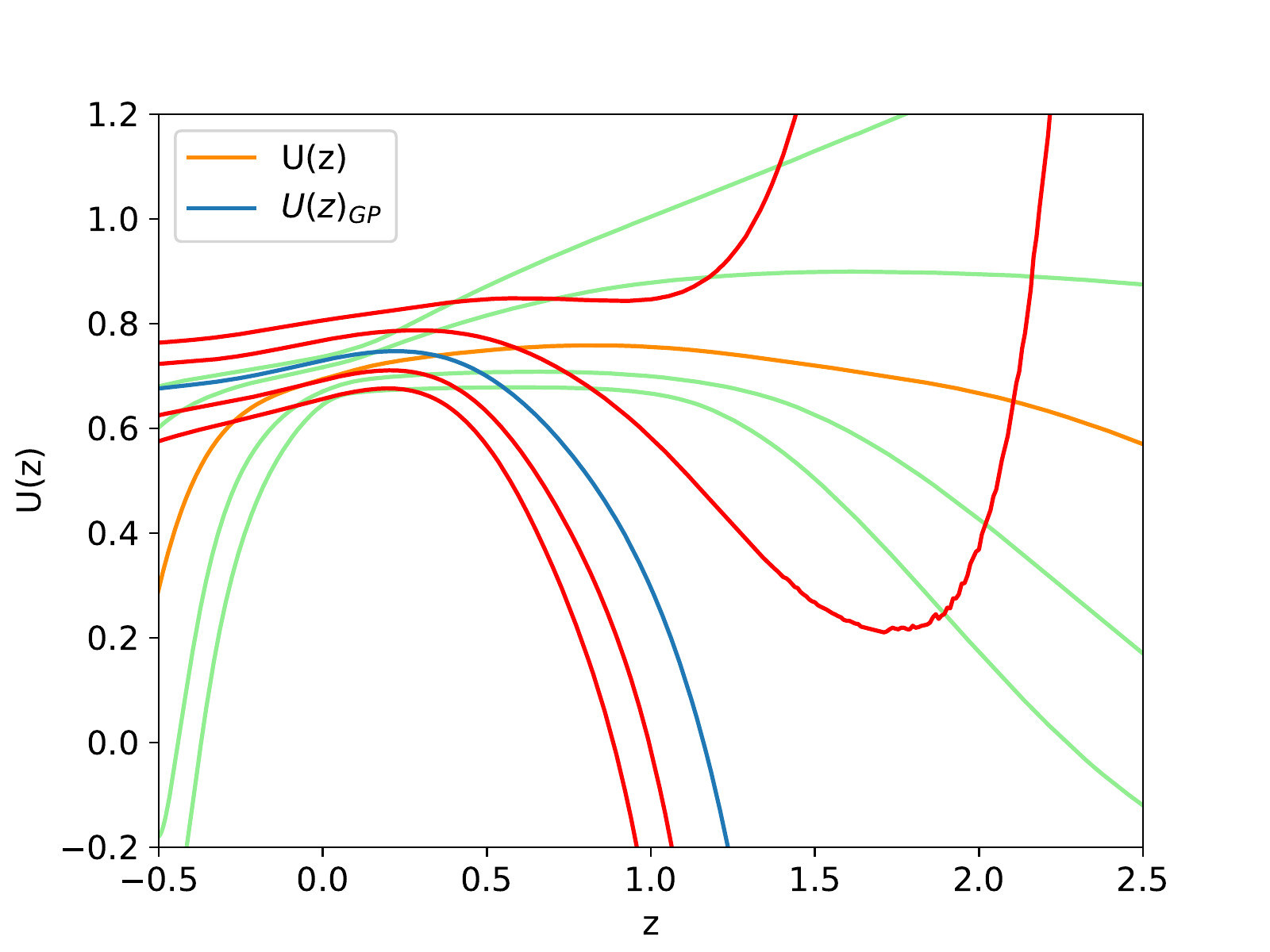}
         \caption{Plot of the $U(z)$ of the analyzed model together with $U(z)$ reconstructed by the GP.  {Also shown are the $1\sigma$ and $2\sigma$ confidence intervals, in red for the model and in green for the GP reconstructions.} \textbf{Left:} GP reconstruction from $H(z)$ data. \textbf{Right:} GP reconstruction from Pantheon.}  
       \label{URec}
\end{figure*}
The reconstruction of the dimensionless potential $U(z)$ obtained by the GP has a downward trend as shown in the Figure \ref{URec}. In the same figure we also show the potential $U(z)$ of the model that has a more constant form. The left figure presents the reconstruction of $U(z)$ for the $H(z)$ data, and  we also show the $U(z)$ of the analysed model. We see the compatibility of the results within $1 \sigma$ or less over the entire redshift range. On the right, we show $U(z)$ reconstructed from the SNe Ia data in contrast to the curve for $U(z)$ of the studied model. We see that both are compatible in $1\sigma$ or less in almost the entire analysed redshift range. 

\section{\label{conc} Conclusions and final comments}

As remarked in the introduction, the possible slowing down of the future cosmic expansion was discussed long ago through a cosmographic approach \cite{Guimaraes:2010mw}. Here,  we have tested  the scalar field plus CDM model as proposed in \cite{Ademir}, which allows for a deceleration of the expansion in the future. In our statistical analysis we have adopted the Pantheon SNe Ia sample and the latest $H(z)$ data.

In the context of this model, we have found a large possibility for deceleration in the distant future, as can be seen on Fig. \ref{qRec}. We have also tried to find evidence for future deceleration in the context of a model independent method, namely, the Gaussian Processes. In this case, by using the $H(z)$ data, we have also obtained a large possibility for deceleration in the vicinity of the future redshift $z\approx-0.5$. However, by using SNe Ia data, no evidence for future deceleration was found in this model-independent method based on Gaussian Processes.

{ It should be stressed that a decelerating regime in the future is impossible not only in the context of the cosmic concordance cosmology ($\Lambda$CDM) but also in the Ratra-Peebles model. However, such a prediction comes out analytically  when the $\alpha$-parameter is added to the original Ratra-Peebles approach. It should also be remarked that de Sitter solution is a  future stable attactor of the $\Lambda$CDM cosmology regardless of the spatial curvature. In this concern, based on qualitative phase space techniques, it seems interesting to investigate whether a similar decelerating attractor exists and how it depends on the values of the $\alpha$ parameter. }

In principle, given the present observed tensions in the $\Lambda$CDM model, it is urgent to verify whether such tensions can be solved or at least alleviated in this enlarged quintessence framework. {It has been shown in Sec. \ref{result} that no considerable difference to $\Lambda$CDM was obtained for $H_0$ with the present data, but future data, as well as a CMB analysis may shed light on the $H_0$ tension in the context of the present model.} In addition, since the model at intermediate redshifts evolves in a slightly different way of $\Lambda$CDM model, and, finally, departs considerably of it, it seems very compelling to investigate all tests related to the theory of small density fluctuations. Further analysis, including different datasets, other non-parametric methods or even the possible influence of the spatial curvature in this framework, will be postpone to a forthcoming communication.

\begin{acknowledgments} SHP acknowledges the National Council for Scientific and Technological Development (CNPq) under grants 303583/2018-5 and 308469/2021-6.  JASL is also  partially supported by CNPQ (310038/2019-7),  CAPES (88881.068485/2014), and FAPESP (LLAMA Project No. 11/51676-9). This study was financed in part by the Coordenação de Aperfeiçoamento de Pessoal de Nível Superior - Brasil (CAPES) - Finance Code 001. The authors would like to thank the two anonymous referees for their constructive suggestions which improved the original version of our paper.
\end{acknowledgments}

\appendix

\section{\label{ap2} Quintessence cosmology with future deceleration}
The authors of reference \cite{Ademir} proposed a class of quintessential cosmological models that allows for a possible deceleration of the universe in the future. Here we will discuss with more detail the expressions presented in that article. It is important to mention that this class of models has the Ratra-Peebles model as a particular case, for $\alpha=0$, and it allows for future deceleration only when $\alpha>0$. For a flat geometry, the Friedmann equation plus the energy conservation expressions for a model driven by a scalar field plus cold dark matter (separately conserved) take the form:

\begin{align}
    H^2 & = \dfrac{8 \pi G}{3}\left(\rho_m + \rho_{\phi}\right),
    \label{eqH2}\\
    \dot{\rho}_m & + 3H\rho_m=0\label{rhom},\\
    \dot{\rho}_{\phi} & + 3H(\rho_{\phi} + p_{\phi}) = 0\label{rhophi}\,
\end{align}
where the energy density and pressure of the scalar field are defined by equations (\ref{eqrho}) and (\ref{eqp}) while its equation of motion is given by (\ref{EM1})

By using equations \eqref{eqrho} and \eqref{eqp}, we see that \eqref{rhophi} becomes:
 \begin{align}\label{phip}
        \dot{\rho}_{\phi}+3H\dot{\phi}^2 &= 0\,.
\end{align}
Now, changing the time derivative for the scale factor $a$ through the identity:
\begin{align}
    \dfrac{d}{dt} = \dfrac{da}{dt}\dfrac{d}{da} = a H  \dfrac{d}{da}\,,
\end{align}
For a universe filled only with field $\phi$, we obtain for \eqref{phip}:
\begin{align}
              \dfrac{d\phi}{da}&=  \sqrt{-\dfrac{1}{a8\pi G }\dfrac{1}{\rho_{\phi}}\dfrac{d\rho_{\phi}}{da}}\,.
              \label{dphida}
\end{align}

Now, let us consider the \textit{ansatz} proposed in \cite{Ademir}:
\begin{align}
    \dfrac{1}{\rho_{\phi}}\dfrac{d\rho_{\phi}}{da} = -\dfrac{\lambda}{a^{1-2\alpha}},
    \label{APADM}
\end{align}
which corresponds to the Ratra-Peebles assumption for $\alpha =0$ \cite{Ratra:1987rm}. In general one finds,
\begin{align}\label{dphi_da}
         d\phi&=  \sqrt{\sigma}a^{-(1-\alpha)}da,
\end{align}
where we have defined
\begin{align}
    \sigma\equiv\dfrac{\lambda}{8\pi G }\,. 
\end{align}
Note also that a simple integration of \eqref{dphida} yields:
\begin{align}
\phi- \phi_0 &= \sqrt{\sigma}\left( \dfrac{a^{\alpha}-1}{\alpha} \right)\,,
\end{align}
so that we can use the relation \eqref{PHiDef} to find the dimensionless field $\Phi$:
\begin{align}\nonumber
    \Phi-\Phi_0 &=\sqrt{\dfrac{8 \pi G}{ 3}}\sqrt{\dfrac{\lambda}{8\pi G }}\left( \dfrac{a^{\alpha}-1}{\alpha} \right)= \sqrt{\dfrac{\lambda}{3}}\left( \dfrac{a^{\alpha}-1}{\alpha} \right)\,.
\end{align}
The expression above allows us to write $a(\Phi)$, as
\begin{align}\label{aalpha}
   a^{\alpha}(\Phi) & =  \alpha\sqrt{\dfrac{3}{\lambda}}\left( \Phi-\Phi_0\right) +1\,.
\end{align}
As we shall see, it is more advantageous to leave $a^{\alpha}(\Phi)$ defined instead of $a(\Phi)$.

Now let us determine a specific form of the potential $V(\Phi)$, which is obtained from \eqref{eqrho}:
\begin{align}\label{V(PHI)}
      V(\Phi)&=\rho_{\phi}\left[1-\dfrac{a^2}{2}\left( \dfrac{d\Phi}{da} \right)^2\right] \,.
\end{align}
Since
\begin{align}
    \dfrac{d\Phi}{da} & =\dfrac{d}{da} \left( \sqrt{\dfrac{\lambda}{3}}\left( \dfrac{a^{\alpha}-1}{\alpha} \right) \right)=
\sqrt{\dfrac{\lambda}{3}}\left( \dfrac{\alpha a^{\alpha-1}}{\alpha} \right) = \sqrt{\dfrac{\lambda}{3}} a^{\alpha-1}\,,
\end{align}
we return to \eqref{V(PHI)} and obtain:
\begin{align}\label{VPHIII}
      V(\Phi)&=\rho_{\phi}\left[1-\dfrac{\lambda}{6} a^{2\alpha} \right]\,.
\end{align}
We can determine $\rho_{\phi}$ according to the \textit{ansatz} \eqref{APADM}, where
\begin{align}\label{rhophiii}
     \rho_{\phi}&= \rho_{\Phi0}e^{-\frac{\lambda}{2\alpha}\left(a^{2\alpha}-1  \right)}\,.
\end{align}
For completeness, we use the expression \eqref{aalpha} to obtain:
\begin{align}\nonumber
    a^{2\alpha} &= \left(  \alpha\sqrt{\dfrac{3}{\lambda}}\left( \Phi-\Phi_0\right) +1\right)^2\,,
    \\\nonumber a^{2\alpha} &=\dfrac{3\alpha^2}{\lambda}\left( \Phi-\Phi_0\right)^2+2\alpha\sqrt{\dfrac{3}{\lambda}}\left( \Phi-\Phi_0\right) +1,
\end{align}
And finally:
\begin{align}\label{rhodphi}
     \rho_{\Phi}&= \rho_{\Phi0}e^{-\frac{1}{2}\left[3\alpha\left( \Phi-\Phi_0\right)^2+2\sqrt{3\lambda}\left( \Phi-\Phi_0 \right)\right]}.
\end{align}

Before writing the final expression for $V(\Phi)$, let's consider an initial condition for the field $\Phi$, given by $\Phi_0 = 0$,  {which is the same initial condition assumed in Ref. \cite{Ademir}}. Then we get for \eqref{VPHIII}:
\begin{align}
    V(\Phi) &= \rho_{\Phi0}e^{-\frac{1}{2}\left[3\alpha \Phi^2+2\sqrt{3\lambda}\Phi\right]}\left[1-\dfrac{\lambda}{6}\left( \alpha\sqrt{\dfrac{3}{\lambda}}\Phi +1\right)^2 \right]\,,
\end{align}
and using the definition \eqref{UDEF}, we find an expression for $U(\Phi)$ given by:
\begin{align}
U(\Phi)   = \Omega_{{\Phi0}}e^{-\frac{1}{2}\left[3\alpha \Phi^2+2\sqrt{3\lambda}\Phi\right]}\left[1-\dfrac{\lambda}{6}\left( \alpha\sqrt{\dfrac{3}{\lambda}}\Phi +1\right)^2 \right]\,,\label{B28}
\end{align}
where we have used $H^2_0 = \dfrac{8 \pi G}{3} \rho_{c0}$ and $\rho_{\Phi0} = \rho_{c0}\Omega_{\Phi0}$.

Now that we have an expression for the dimensionless potential $U(\Phi)$, we can rewrite the equation for $E^2$ given by
\begin{align}\label{EPHI}
    E^2 &= \dfrac{\Omega_{m0}e^{-3N}+  \Omega_{{\Phi0}}e^{-\frac{1}{2}\left[3\alpha \Phi^2+2\sqrt{3\lambda}\Phi\right]}\left[1-\dfrac{\lambda}{6}\left( \alpha\sqrt{\dfrac{3}{\lambda}}\Phi +1\right)^2 \right]}{1-\dfrac{\Phi'^2}{2}}.
\end{align}

For the equation of motion of the field $\Phi$, we have:
\begin{widetext}
\begin{align}
      \Phi'' +\dfrac{1}{2E^2} \left\{\left(3\Omega_{m0}e^{-3N}+6\Omega_{{\Phi0}}e^{-\frac{1}{2}\left[3\alpha \Phi^2+2\sqrt{3\lambda}\Phi\right]}\left[1-\dfrac{\lambda}{6}\left( \alpha\sqrt{\dfrac{3}{\lambda}}\Phi +1\right)^2 \right]\right)\Phi' +2\frac{dU}{d\Phi}\right\}= 0\,.\label{eqphiN}
\end{align}
where 
\begin{align}\nonumber
 \frac{dU}{d\Phi}= \frac{1}{6} \Omega_{\Phi0} e^{-\frac{1}{2} \left[3 \alpha \Phi^2+2 \sqrt{3\lambda}\Phi\right]}\left\{3 \alpha \Phi \left[\alpha \left(3 \alpha \Phi^2+3\sqrt{3\lambda}\Phi-2\right)+3\lambda-6\right]+\sqrt{3\lambda} (\lambda-6-2\alpha)\right\}\,.
\end{align}
\end{widetext}
It is important to mention that in Eqs. \eqref{EPHI} and \eqref{eqphiN}, the primes correspond to derivatives with respect to $N\equiv \ln(a)$, the number of e-folds. 
\nocite{*}


\begin{thebibliography}{47}%
\makeatletter
\providecommand \@ifxundefined [1]{%
 \@ifx{#1\undefined}
}%
\providecommand \@ifnum [1]{%
 \ifnum #1\expandafter \@firstoftwo
 \else \expandafter \@secondoftwo
 \fi
}%
\providecommand \@ifx [1]{%
 \ifx #1\expandafter \@firstoftwo
 \else \expandafter \@secondoftwo
 \fi
}%
\providecommand \natexlab [1]{#1}%
\providecommand \enquote  [1]{``#1''}%
\providecommand \bibnamefont  [1]{#1}%
\providecommand \bibfnamefont [1]{#1}%
\providecommand \citenamefont [1]{#1}%
\providecommand \href@noop [0]{\@secondoftwo}%
\providecommand \href [0]{\begingroup \@sanitize@url \@href}%
\providecommand \@href[1]{\@@startlink{#1}\@@href}%
\providecommand \@@href[1]{\endgroup#1\@@endlink}%
\providecommand \@sanitize@url [0]{\catcode `\\12\catcode `\$12\catcode
  `\&12\catcode `\#12\catcode `\^12\catcode `\_12\catcode `\%12\relax}%
\providecommand \@@startlink[1]{}%
\providecommand \@@endlink[0]{}%
\providecommand \url  [0]{\begingroup\@sanitize@url \@url }%
\providecommand \@url [1]{\endgroup\@href {#1}{\urlprefix }}%
\providecommand \urlprefix  [0]{URL }%
\providecommand \Eprint [0]{\href }%
\providecommand \doibase [0]{https://doi.org/}%
\providecommand \selectlanguage [0]{\@gobble}%
\providecommand \bibinfo  [0]{\@secondoftwo}%
\providecommand \bibfield  [0]{\@secondoftwo}%
\providecommand \translation [1]{[#1]}%
\providecommand \BibitemOpen [0]{}%
\providecommand \bibitemStop [0]{}%
\providecommand \bibitemNoStop [0]{.\EOS\space}%
\providecommand \EOS [0]{\spacefactor3000\relax}%
\providecommand \BibitemShut  [1]{\csname bibitem#1\endcsname}%
\let\auto@bib@innerbib\@empty


\bibitem [{\citenamefont {Riess}\ \emph {et~al.}(1998)\citenamefont {Riess}
  \emph {et~al.}}]{SN1}%
  \BibitemOpen
  \bibfield  {author} {\bibinfo {author} {\bibfnamefont {A.~G.}\ \bibnamefont
  {Riess}} \emph {et~al.} (\bibinfo {collaboration} {Supernova Search Team}),\
  }\href@noop {} {\bibfield  {journal} {\bibinfo  {journal} {Astron. J.}\
  }\textbf {\bibinfo {volume} {116}},\ \bibinfo {pages} {1009} (\bibinfo {year}
  {1998})}\BibitemShut {NoStop}%
\bibitem [{\citenamefont {Perlmutter}\ \emph {et~al.}(1999)\citenamefont
  {Perlmutter} \emph {et~al.}}]{SN2}%
  \BibitemOpen
  \bibfield  {author} {\bibinfo {author} {\bibfnamefont {S.}~\bibnamefont
  {Perlmutter}} \emph {et~al.} (\bibinfo {collaboration} {Supernova Cosmology
  Project}),\ }\href@noop {} {\bibfield  {journal} {\bibinfo  {journal}
  {Astrophys. J.}\ }\textbf {\bibinfo {volume} {517}},\ \bibinfo {pages} {565}
  (\bibinfo {year} {1999})}

\bibitem [{\citenamefont {Aghanim}\ \emph {et~al.}(2020)\citenamefont {Aghanim}
  \emph {et~al.}}]{Planck2018}%
  \BibitemOpen
  \bibfield  {author} {\bibinfo {author} {\bibfnamefont {N.}~\bibnamefont
  {Aghanim}} \emph {et~al.} (\bibinfo {collaboration} {Planck}),\ }\href@noop
  {} {\bibfield  {journal} {\bibinfo  {journal} {Astron. Astrophys.}\ }\textbf
  {\bibinfo {volume} {641}},\ \bibinfo {pages} {A6} (\bibinfo {year} {2020})},\
  \bibinfo {note} {[Erratum: Astron.Astrophys. 652, C4 (2021)]}


\bibitem{BBN}
R.~L.~Workman \textit{et al.} [Particle Data Group],
PTEP \textbf{2022} (2022), 083C01, Ch. 24


\bibitem{SDSS:2005xqv}
D.~J.~Eisenstein \textit{et al.} [SDSS],
Astrophys. J. \textbf{633}, 560-574 (2005)
[arXiv:astro-ph/0501171 [astro-ph]].

\bibitem{ValentinoH0}
E.~Di Valentino, O.~Mena, S.~Pan, L.~Visinelli, W.~Yang, A.~Melchiorri, D.~F.~Mota, A.~G.~Riess and J.~Silk,
Class. Quant. Grav. \textbf{38} (2021) no.15, 153001
[arXiv:2103.01183 [astro-ph.CO]].

\bibitem{ValentinoS8}
E.~Di Valentino, L.~A.~Anchordoqui, \"O.~Akarsu, Y.~Ali-Haimoud, L.~Amendola, N.~Arendse, M.~Asgari, M.~Ballardini, S.~Basilakos and E.~Battistelli, \textit{et al.}
Astropart. Phys. \textbf{131} (2021), 102604
[arXiv:2008.11285 [astro-ph.CO]].
\bibitem{NearbySupernovaFactory:2018qkd}
M.~Rigault \textit{et al.} [Nearby Supernova Factory],
Astron. Astrophys. \textbf{644} (2020), A176
doi:10.1051/0004-6361/201730404
[arXiv:1806.03849 [astro-ph.CO]].
\bibitem{Riess:2018kzi}
A.~G.~Riess, S.~Casertano, D.~Kenworthy, D.~Scolnic and L.~Macri,
[arXiv:1810.03526 [astro-ph.CO]].

\bibitem [{\citenamefont {Martinelli}\ and\ \citenamefont
  {Tutusaus}(2019)}]{Martinelli:2019krf}%
  \BibitemOpen
  \bibfield  {author} {\bibinfo {author} {\bibfnamefont {M.}~\bibnamefont
  {Martinelli}}\ and\ \bibinfo {author} {\bibfnamefont {I.}~\bibnamefont
  {Tutusaus}},\ }\bibfield  {journal} {\bibinfo  {journal} {Symmetry}\ }\textbf
  {\bibinfo {volume} {11}},\ \href@noop {} {} (\bibinfo {year}
  {2019})\BibitemShut {NoStop}%


\bibitem [{\citenamefont {Del~Popolo}\ \emph {et~al.}(2014)\citenamefont
  {Del~Popolo}, \citenamefont {Lima}, \citenamefont {Fabris},\ and\
  \citenamefont {Rodrigues}}]{P2014}%
  \BibitemOpen
  \bibfield  {author} {\bibinfo {author} {\bibfnamefont {A.}~\bibnamefont
  {Del~Popolo}}, \bibinfo {author} {\bibfnamefont {J.~A.~S.}\ \bibnamefont
  {Lima}}, \bibinfo {author} {\bibfnamefont {J.~C.}\ \bibnamefont {Fabris}},\
  and\ \bibinfo {author} {\bibfnamefont {D.~C.}\ \bibnamefont {Rodrigues}},\
  }\href@noop {} {\bibfield  {journal} {\bibinfo  {journal} {JCAP}\ }\textbf
  {\bibinfo {volume} {2014}}\bibinfo  {number} { (04)},\ \bibinfo {pages}
  {021}}\BibitemShut {NoStop}%
  \,[arXiv:1404.3674 [astro-ph.CO]]
\bibitem [{\citenamefont {Riess}(2019)}]{Riess:2020sih}%
  \BibitemOpen
\bibfield  {number} {  }\bibfield  {author} {\bibinfo {author} {\bibfnamefont
  {A.~G.}\ \bibnamefont {Riess}},\ }\href@noop {} {\bibfield  {journal}
  {\bibinfo  {journal} {Nature Rev. Phys.}\ }\textbf {\bibinfo {volume} {2}},\
  \bibinfo {pages} {10} (\bibinfo {year} {2019})}\BibitemShut {NoStop}%

  
\bibitem [{\citenamefont {Bull}\ \emph {et~al.}(2016)\citenamefont {Bull} \emph
  {et~al.}}]{Bull:2015stt}%
  \BibitemOpen
  \bibfield  {author} {\bibinfo {author} {\bibfnamefont {P.}~\bibnamefont
  {Bull}} \emph {et~al.},\ }\href@noop {} {\bibfield  {journal} {\bibinfo
  {journal} {Phys. Dark Univ.}\ }\textbf {\bibinfo {volume} {12}},\ \bibinfo
  {pages} {56} (\bibinfo {year} {2016})}\BibitemShut {NoStop}%

\bibitem [{\citenamefont {Martin}(2008)}]{Martin:2008qp}%
  \BibitemOpen
  \bibfield  {author} {\bibinfo {author} {\bibfnamefont {J.}~\bibnamefont
  {Martin}},\ }\href@noop {} {\bibfield  {journal} {\bibinfo  {journal} {Mod.
  Phys. Lett. A}\ }\textbf {\bibinfo {volume} {23}},\ \bibinfo {pages} {1252}
  (\bibinfo {year} {2008})}\BibitemShut {NoStop}%
\bibitem [{\citenamefont {Tsujikawa}(2013)}]{Tsujikawa:2013fta}%
  \BibitemOpen
  \bibfield  {author} {\bibinfo {author} {\bibfnamefont {S.}~\bibnamefont
  {Tsujikawa}},\ }\href@noop {} {\bibfield  {journal} {\bibinfo  {journal}
  {Class. Quant. Grav.}\ }\textbf {\bibinfo {volume} {30}},\ \bibinfo {pages}
  {214003} (\bibinfo {year} {2013})}\BibitemShut {NoStop}%
\bibitem [{\citenamefont {Guth}(1981)}]{GUT}%
  \BibitemOpen
  \bibfield  {author} {\bibinfo {author} {\bibfnamefont {A.~H.}\ \bibnamefont
  {Guth}},\ }\href@noop {} {\bibfield  {journal} {\bibinfo  {journal} {Phys.
  Rev. D}\ }\textbf {\bibinfo {volume} {23}},\ \bibinfo {pages} {347} (\bibinfo
  {year} {1981})}\BibitemShut {NoStop}%
\bibitem [{\citenamefont {Liddle}\ and\ \citenamefont {Lyth}(2000)}]{LCF}%
  \BibitemOpen
  \bibfield  {author} {\bibinfo {author} {\bibfnamefont {A.~R.}\ \bibnamefont
  {Liddle}}\ and\ \bibinfo {author} {\bibfnamefont {D.~H.}\ \bibnamefont
  {Lyth}},\ }\href@noop {} {\emph {\bibinfo {title} {{Cosmological inflation
  and large scale structure}}}}\ (\bibinfo  {publisher} {2000},\ \bibinfo
  {year} {2000})\BibitemShut {NoStop}%
\bibitem [{\citenamefont {S\'a}(2020)}]{Sa:2020fvn}%
  \BibitemOpen
  \bibfield  {author} {\bibinfo {author} {\bibfnamefont {P.~M.}\ \bibnamefont
  {S\'a}},\ }\href@noop {} {\bibfield  {journal} {\bibinfo  {journal} {Phys.
  Rev. D}\ }\textbf {\bibinfo {volume} {102}},\ \bibinfo {pages} {103519}
  (\bibinfo {year} {2020})}\BibitemShut {NoStop}%
\bibitem [{\citenamefont {Lin}(2009)}]{Lin:2009ta}%
  \BibitemOpen
  \bibfield  {author} {\bibinfo {author} {\bibfnamefont {Chia-Min}\ \bibnamefont
  {Lin}}\ }\href@noop {} {\bibfield  {journal} {\bibinfo  {journal} {arXiv:0906.5021 [hep-ph]}\ }
  (\bibinfo {year} {2009})}\BibitemShut {NoStop}%
\bibitem [{\citenamefont {Liddle}\ \emph {et~al.}(2008)\citenamefont {Liddle},
  \citenamefont {Pahud},\ and\ \citenamefont {Urena-Lopez}}]{Liddle:2008bm}%
  \BibitemOpen
  \bibfield  {author} {\bibinfo {author} {\bibfnamefont {A.~R.}\ \bibnamefont
  {Liddle}}, \bibinfo {author} {\bibfnamefont {C.}~\bibnamefont {Pahud}},\ and\
  \bibinfo {author} {\bibfnamefont {L.~A.}\ \bibnamefont {Urena-Lopez}},\
  }\href@noop {} {\bibfield  {journal} {\bibinfo  {journal} {Phys. Rev. D}\
  }\textbf {\bibinfo {volume} {77}},\ \bibinfo {pages} {121301} (\bibinfo
  {year} {2008})}\BibitemShut {NoStop}%
\bibitem [{\citenamefont {Escobal}\ \emph {et~al.}(2021)\citenamefont
  {Escobal}, \citenamefont {Jesus},\ and\ \citenamefont {Pereira}}]{Escobal}%
  \BibitemOpen
  \bibfield  {author} {\bibinfo {author} {\bibfnamefont {A.~A.}\ \bibnamefont
  {Escobal}}, \bibinfo {author} {\bibfnamefont {J.~F.}\ \bibnamefont {Jesus}},\
  and\ \bibinfo {author} {\bibfnamefont {S.~H.}\ \bibnamefont {Pereira}},\
  }\href@noop {} {\bibfield  {journal} {\bibinfo  {journal} {Int. J. Mod. Phys.
  D}\ }\textbf {\bibinfo {volume} {30}},\ \bibinfo {pages} {2150108} (\bibinfo
  {year} {2021})}\BibitemShut {NoStop}%
\bibitem [{\citenamefont {Maga\~na}\ and\ \citenamefont
  {Matos}(2012)}]{Magana012}%
  \BibitemOpen
  \bibfield  {author} {\bibinfo {author} {\bibfnamefont {J.}~\bibnamefont
  {Maga\~na}}\ and\ \bibinfo {author} {\bibfnamefont {T.}~\bibnamefont
  {Matos}},\ }\href@noop {} {\bibfield  {journal} {\bibinfo  {journal} {J.
  Phys. Conf. Ser.}\ }\textbf {\bibinfo {volume} {378}},\ \bibinfo {pages}
  {012012} (\bibinfo {year} {2012})}\BibitemShut {NoStop}%
\bibitem [{\citenamefont {Sahni}\ and\ \citenamefont
  {Wang}(2000)}]{Sahni:1999qe}%
  \BibitemOpen
  \bibfield  {author} {\bibinfo {author} {\bibfnamefont {V.}~\bibnamefont
  {Sahni}}\ and\ \bibinfo {author} {\bibfnamefont {L.-M.}\ \bibnamefont
  {Wang}},\ }\href@noop {} {\bibfield  {journal} {\bibinfo  {journal} {Phys.
  Rev. D}\ }\textbf {\bibinfo {volume} {62}},\ \bibinfo {pages} {103517}
  (\bibinfo {year} {2000})}\BibitemShut {NoStop}%
\bibitem [{\citenamefont {Rubano}\ and\ \citenamefont
  {Barrow}(2001)}]{Rubano:2001xi}%
  \BibitemOpen
  \bibfield  {author} {\bibinfo {author} {\bibfnamefont {C.}~\bibnamefont
  {Rubano}}\ and\ \bibinfo {author} {\bibfnamefont {J.~D.}\ \bibnamefont
  {Barrow}},\ }\href@noop {} {\bibfield  {journal} {\bibinfo  {journal} {Phys.
  Rev. D}\ }\textbf {\bibinfo {volume} {64}},\ \bibinfo {pages} {127301}
  (\bibinfo {year} {2001})}\BibitemShut {NoStop}%
\bibitem [{\citenamefont {Paliathanasis}\ \emph {et~al.}(2014)\citenamefont
  {Paliathanasis}, \citenamefont {Tsamparlis},\ and\ \citenamefont
  {Basilakos}}]{Paliathanasis:2014zxa}%
  \BibitemOpen
  \bibfield  {author} {\bibinfo {author} {\bibfnamefont {A.}~\bibnamefont
  {Paliathanasis}}, \bibinfo {author} {\bibfnamefont {M.}~\bibnamefont
  {Tsamparlis}},\ and\ \bibinfo {author} {\bibfnamefont {S.}~\bibnamefont
  {Basilakos}},\ }\href@noop {} {\bibfield  {journal} {\bibinfo  {journal}
  {Phys. Rev. D}\ }\textbf {\bibinfo {volume} {90}},\ \bibinfo {pages} {103524}
  (\bibinfo {year} {2014})}\BibitemShut {NoStop}%
\bibitem [{\citenamefont {Dimakis}\ \emph {et~al.}(2016)\citenamefont
  {Dimakis}, \citenamefont {Karagiorgos}, \citenamefont {Zampeli},
  \citenamefont {Paliathanasis}, \citenamefont {Christodoulakis},\ and\
  \citenamefont {Terzis}}]{Dimakis:2016mip}%
  \BibitemOpen
  \bibfield  {author} {\bibinfo {author} {\bibfnamefont {N.}~\bibnamefont
  {Dimakis}}, \bibinfo {author} {\bibfnamefont {A.}~\bibnamefont
  {Karagiorgos}}, \bibinfo {author} {\bibfnamefont {A.}~\bibnamefont
  {Zampeli}}, \bibinfo {author} {\bibfnamefont {A.}~\bibnamefont
  {Paliathanasis}}, \bibinfo {author} {\bibfnamefont {T.}~\bibnamefont
  {Christodoulakis}},\ and\ \bibinfo {author} {\bibfnamefont {P.~A.}\
  \bibnamefont {Terzis}},\ }\href@noop {} {\bibfield  {journal} {\bibinfo
  {journal} {Phys. Rev. D}\ }\textbf {\bibinfo {volume} {93}},\ \bibinfo
  {pages} {123518} (\bibinfo {year} {2016})}\BibitemShut {NoStop}%
\bibitem [{\citenamefont {Ratra}(1991)}]{Ratra:1990me}%
  \BibitemOpen
  \bibfield  {author} {\bibinfo {author} {\bibfnamefont {B.}~\bibnamefont
  {Ratra}},\ }\href@noop {} {\bibfield  {journal} {\bibinfo  {journal} {Phys.
  Rev. D}\ }\textbf {\bibinfo {volume} {44}},\ \bibinfo {pages} {352} (\bibinfo
  {year} {1991})}\BibitemShut {NoStop}%
\bibitem [{\citenamefont {Urena-Lopez}\ and\ \citenamefont
  {Reyes-Ibarra}(2009)}]{Urena-Lopez:2007zal}%
  \BibitemOpen
  \bibfield  {author} {\bibinfo {author} {\bibfnamefont {L.~A.}\ \bibnamefont
  {Urena-Lopez}}\ and\ \bibinfo {author} {\bibfnamefont {M.~J.}\ \bibnamefont
  {Reyes-Ibarra}},\ }\href@noop {} {\bibfield  {journal} {\bibinfo  {journal}
  {Int. J. Mod. Phys. D}\ }\textbf {\bibinfo {volume} {18}},\ \bibinfo {pages}
  {621} (\bibinfo {year} {2009})}\BibitemShut {NoStop}%
\bibitem [{\citenamefont {Peebles}\ and\ \citenamefont
  {Ratra}(1988)}]{Peebles:1987ek}%
  \BibitemOpen
  \bibfield  {author} {\bibinfo {author} {\bibfnamefont {P.~J.~E.}\
  \bibnamefont {Peebles}}\ and\ \bibinfo {author} {\bibfnamefont
  {B.}~\bibnamefont {Ratra}},\ }\href@noop {} {\bibfield  {journal} {\bibinfo
  {journal} {Astrophys. J. Lett.}\ }\textbf {\bibinfo {volume} {325}},\
  \bibinfo {pages} {L17} (\bibinfo {year} {1988})}

  \bibitem [{\citenamefont {Ratra}\ and\ \citenamefont
  {Peebles}(1988)}]{Ratra:1987rm}%
  \BibitemOpen
  \bibfield  {author} {\bibinfo {author} {\bibfnamefont {B.}~\bibnamefont
  {Ratra}}\ and\ \bibinfo {author} {\bibfnamefont {P.~J.~E.}\ \bibnamefont
  {Peebles}},\ }\href {https://doi.org/10.1103/PhysRevD.37.3406} {\bibfield
  {journal} {\bibinfo  {journal} {Phys. Rev. D}\ }\textbf {\bibinfo {volume}
  {37}},\ \bibinfo {pages} {3406} (\bibinfo {year} {1988})}\BibitemShut
  {NoStop}%
\bibitem [{\citenamefont {Yang}\ \emph {et~al.}(2019)\citenamefont {Yang},
  \citenamefont {Shahalam}, \citenamefont {Pal}, \citenamefont {Pan},\ and\
  \citenamefont {Wang}}]{Yang:2018xah}%
  \BibitemOpen
  \bibfield  {author} {\bibinfo {author} {\bibfnamefont {W.}~\bibnamefont
  {Yang}}, \bibinfo {author} {\bibfnamefont {M.}~\bibnamefont {Shahalam}},
  \bibinfo {author} {\bibfnamefont {B.}~\bibnamefont {Pal}}, \bibinfo {author}
  {\bibfnamefont {S.}~\bibnamefont {Pan}},\ and\ \bibinfo {author}
  {\bibfnamefont {A.}~\bibnamefont {Wang}},\ }\href@noop {} {\bibfield
  {journal} {\bibinfo  {journal} {Phys. Rev. D}\ }\textbf {\bibinfo {volume}
  {100}},\ \bibinfo {pages} {023522} (\bibinfo {year} {2019})}\BibitemShut
  {NoStop}%
\bibitem [{\citenamefont {Basilakos}\ \emph {et~al.}(2013)\citenamefont
  {Basilakos}, \citenamefont {Lima},\ and\ \citenamefont
  {Sola}}]{Basilakos:2013xpa}%
  \BibitemOpen
  \bibfield  {author} {\bibinfo {author} {\bibfnamefont {S.}~\bibnamefont
  {Basilakos}}, \bibinfo {author} {\bibfnamefont {J.~A.~S.}\ \bibnamefont
  {Lima}},\ and\ \bibinfo {author} {\bibfnamefont {J.}~\bibnamefont {Sola}},\
  }\href@noop {} {\bibfield  {journal} {\bibinfo  {journal} {Int. J. Mod. Phys.
  D}\ }\textbf {\bibinfo {volume} {22}},\ \bibinfo {pages} {1342008} (\bibinfo
  {year} {2013})}\BibitemShut {NoStop}%
\bibitem [{\citenamefont {Lima}\ \emph {et~al.}(2012)\citenamefont {Lima},
  \citenamefont {Basilakos},\ and\ \citenamefont {Costa}}]{Lima:2012cm}%
  \BibitemOpen
  \bibfield  {author} {\bibinfo {author} {\bibfnamefont {J.~A.~S.}\
  \bibnamefont {Lima}}, \bibinfo {author} {\bibfnamefont {S.}~\bibnamefont
  {Basilakos}},\ and\ \bibinfo {author} {\bibfnamefont {F.~E.~M.}\ \bibnamefont
  {Costa}},\ }\href {https://doi.org/10.1103/PhysRevD.86.103534} {\bibfield
  {journal} {\bibinfo  {journal} {Phys. Rev. D}\ }\textbf {\bibinfo {volume}
  {86}},\ \bibinfo {pages} {103534} (\bibinfo {year} {2012})},\ \Eprint
  {https://arxiv.org/abs/1205.0868} {arXiv:1205.0868 [astro-ph.CO]}
  \BibitemShut {NoStop}%
\bibitem [{\citenamefont {Maia}\ and\ \citenamefont
  {Lima}(2002)}]{Maia:2001zu}%
  \BibitemOpen
  \bibfield  {author} {\bibinfo {author} {\bibfnamefont {J.~M.~F.}\
  \bibnamefont {Maia}}\ and\ \bibinfo {author} {\bibfnamefont {J.~A.~S.}\
  \bibnamefont {Lima}},\ }\href@noop {} {\bibfield  {journal} {\bibinfo
  {journal} {Phys. Rev. D}\ }\textbf {\bibinfo {volume} {65}},\ \bibinfo
  {pages} {083513} (\bibinfo {year} {2002})}

  \bibitem [{\citenamefont {Carvalho}\ \emph {et~al.}(2006)\citenamefont
  {Carvalho}, \citenamefont {Alcaniz}, \citenamefont {Lima},\ and\
  \citenamefont {Silva}}]{Ademir}%
  \BibitemOpen
  \bibfield  {author} {\bibinfo {author} {\bibfnamefont {F.~C.}\ \bibnamefont
  {Carvalho}}, \bibinfo {author} {\bibfnamefont {J.~S.}\ \bibnamefont
  {Alcaniz}}, \bibinfo {author} {\bibfnamefont {J.~A.~S.}\ \bibnamefont
  {Lima}},\ and\ \bibinfo {author} {\bibfnamefont {R.}~\bibnamefont {Silva}},\
  }\href@noop {} {\bibfield  {journal} {\bibinfo  {journal} {Phys. Rev. Lett.}\
  }\textbf {\bibinfo {volume} {97}},\ \bibinfo {pages} {081301} (\bibinfo
  {year} {2006})}\BibitemShut {NoStop}%
\bibitem [{\citenamefont {Castillo-Santos}\ \emph {et~al.}(2023)\citenamefont
  {Castillo-Santos}, \citenamefont {Hern\'andez-Almada}, \citenamefont
  {Garc\'\i{}a-Aspeitia},\ and\ \citenamefont {Maga\~na}}]{PDU2023}%
  \BibitemOpen
  \bibfield  {author} {\bibinfo {author} {\bibfnamefont {M.~N.}\ \bibnamefont
  {Castillo-Santos}}, \bibinfo {author} {\bibfnamefont {A.}~\bibnamefont
  {Hern\'andez-Almada}}, \bibinfo {author} {\bibfnamefont {M.~A.}\ \bibnamefont
  {Garc\'\i{}a-Aspeitia}},\ and\ \bibinfo {author} {\bibfnamefont
  {J.}~\bibnamefont {Maga\~na}},\ }\href@noop {} {\bibfield  {journal}
  {\bibinfo  {journal} {Phys. Dark Univ.}\ }\textbf {\bibinfo {volume} {40}},\
  \bibinfo {pages} {101225} (\bibinfo {year} {2023})}\BibitemShut {NoStop}%
\bibitem [{\citenamefont {Fischler}\ \emph {et~al.}(2001)\citenamefont
  {Fischler}, \citenamefont {Kashani-Poor}, \citenamefont {McNees},\ and\
  \citenamefont {Paban}}]{Fischler:2001yj}%
  \BibitemOpen
  \bibfield  {author} {\bibinfo {author} {\bibfnamefont {W.}~\bibnamefont
  {Fischler}}, \bibinfo {author} {\bibfnamefont {A.}~\bibnamefont
  {Kashani-Poor}}, \bibinfo {author} {\bibfnamefont {R.}~\bibnamefont
  {McNees}},\ and\ \bibinfo {author} {\bibfnamefont {S.}~\bibnamefont
  {Paban}},\ }\href@noop {} {\bibfield  {journal} {\bibinfo  {journal} {JHEP}\
  }\textbf {\bibinfo {volume} {2001}}\bibinfo  {number} { (07)},\ \bibinfo
  {pages} {003}}\BibitemShut {NoStop}%
\bibitem [{\citenamefont {Cline}(2001)}]{Cline:2001nq}%
  \BibitemOpen
\bibfield  {number} {  }\bibfield  {author} {\bibinfo {author} {\bibfnamefont
  {J.~M.}\ \bibnamefont {Cline}},\ }\href@noop {} {\bibfield  {journal}
  {\bibinfo  {journal} {JHEP}\ }\textbf {\bibinfo {volume} {2021}}\bibinfo
  {number} { (08)},\ \bibinfo {pages} {035}}\BibitemShut {NoStop}%
\bibitem [{\citenamefont {Scolnic}\ \emph {et~al.}(2018)\citenamefont {Scolnic}
  \emph {et~al.}}]{Scolnic:2017caz}%
  \BibitemOpen
\bibfield  {number} {  }\bibfield  {author} {\bibinfo {author} {\bibfnamefont
  {D.~M.}\ \bibnamefont {Scolnic}} \emph {et~al.} (\bibinfo {collaboration}
  {Pan-STARRS1}),\ }\href@noop {} {\bibfield  {journal} {\bibinfo  {journal}
  {Astrophys. J.}\ }\textbf {\bibinfo {volume} {859}},\ \bibinfo {pages} {101}
  (\bibinfo {year} {2018})}\BibitemShut {NoStop}%
\bibitem [{\citenamefont {Maga\~na}\ \emph {et~al.}(2018)\citenamefont
  {Maga\~na}, \citenamefont {Amante}, \citenamefont {Garcia-Aspeitia},\ and\
  \citenamefont {Motta}}]{Magana:2017nfs}%
  \BibitemOpen
  \bibfield  {author} {\bibinfo {author} {\bibfnamefont {J.}~\bibnamefont
  {Maga\~na}}, \bibinfo {author} {\bibfnamefont {M.~H.}\ \bibnamefont
  {Amante}}, \bibinfo {author} {\bibfnamefont {M.~A.}\ \bibnamefont
  {Garcia-Aspeitia}},\ and\ \bibinfo {author} {\bibfnamefont {V.}~\bibnamefont
  {Motta}},\ }\href@noop {} {\bibfield  {journal} {\bibinfo  {journal} {Mon.
  Not. Roy. Astron. Soc.}\ }\textbf {\bibinfo {volume} {476}},\ \bibinfo
  {pages} {1036} (\bibinfo {year} {2018})}\BibitemShut {NoStop}%
\bibitem [{\citenamefont {Guimar\~aes}\ and\ \citenamefont
  {Lima}(2011)}]{Guimaraes:2010mw}%
  \BibitemOpen
  \bibfield  {author} {\bibinfo {author} {\bibfnamefont {A.~C.~C.}\
  \bibnamefont {Guimar\~aes}}\ and\ \bibinfo {author} {\bibfnamefont
  {J.~A.~S.}\ \bibnamefont {Lima}},\ }\href@noop {} {\bibfield  {journal}
  {\bibinfo  {journal} {Class. Quant. Grav.}\ }\textbf {\bibinfo {volume}
  {28}},\ \bibinfo {pages} {125026} (\bibinfo {year} {2011})}\BibitemShut
  {NoStop}%
\bibitem [{\citenamefont {Seikel}\ \emph {et~al.}(2012)\citenamefont {Seikel},
  \citenamefont {Clarkson},\ and\ \citenamefont {Smith}}]{Seikel:2012uu}%
  \BibitemOpen
  \bibfield  {author} {\bibinfo {author} {\bibfnamefont {M.}~\bibnamefont
  {Seikel}}, \bibinfo {author} {\bibfnamefont {C.}~\bibnamefont {Clarkson}},\
  and\ \bibinfo {author} {\bibfnamefont {M.}~\bibnamefont {Smith}},\
  }\href@noop {} {\bibfield  {journal} {\bibinfo  {journal} {JCAP}\ }\textbf
  {\bibinfo {volume} {2012}}\bibinfo  {number} { (06)},\ \bibinfo {pages}
  {036}}\BibitemShut {NoStop}%
\bibitem [{\citenamefont {Holsclaw}\ \emph {et~al.}(2010)\citenamefont
  {Holsclaw}, \citenamefont {Alam}, \citenamefont {Sanso}, \citenamefont {Lee},
  \citenamefont {Heitmann}, \citenamefont {Habib},\ and\ \citenamefont
  {Higdon}}]{Holsclaw:2010nb}%
  \BibitemOpen
\bibfield  {number} {  }\bibfield  {author} {\bibinfo {author} {\bibfnamefont
  {T.}~\bibnamefont {Holsclaw}}, \bibinfo {author} {\bibfnamefont
  {U.}~\bibnamefont {Alam}}, \bibinfo {author} {\bibfnamefont {B.}~\bibnamefont
  {Sanso}}, \bibinfo {author} {\bibfnamefont {H.}~\bibnamefont {Lee}}, \bibinfo
  {author} {\bibfnamefont {K.}~\bibnamefont {Heitmann}}, \bibinfo {author}
  {\bibfnamefont {S.}~\bibnamefont {Habib}},\ and\ \bibinfo {author}
  {\bibfnamefont {D.}~\bibnamefont {Higdon}},\ }\href@noop {} {\bibfield
  {journal} {\bibinfo  {journal} {Phys. Rev. D}\ }\textbf {\bibinfo {volume}
  {82}},\ \bibinfo {pages} {103502} (\bibinfo {year} {2010})}\BibitemShut
  {NoStop}%
\bibitem [{\citenamefont {Shafieloo}\ \emph {et~al.}(2012)\citenamefont
  {Shafieloo}, \citenamefont {Kim},\ and\ \citenamefont
  {Linder}}]{Shafieloo:2012ht}%
  \BibitemOpen
  \bibfield  {author} {\bibinfo {author} {\bibfnamefont {A.}~\bibnamefont
  {Shafieloo}}, \bibinfo {author} {\bibfnamefont {A.~G.}\ \bibnamefont {Kim}},\
  and\ \bibinfo {author} {\bibfnamefont {E.~V.}\ \bibnamefont {Linder}},\
  }\href@noop {} {\bibfield  {journal} {\bibinfo  {journal} {Phys. Rev. D}\
  }\textbf {\bibinfo {volume} {85}},\ \bibinfo {pages} {123530} (\bibinfo
  {year} {2012})}

\bibitem{Jesus:2019nnk}
J.~F.~Jesus, R.~Valentim, A.~A.~Escobal and S.~H.~Pereira,
JCAP \textbf{04}, 053 (2020)

\bibitem{MartinEtAl13}
J.~Martin, C.~Ringeval and V.~Vennin,
Phys. Dark Univ. \textbf{5-6} (2014), 75-235
[arXiv:1303.3787 [astro-ph.CO]].

\bibitem{Tripp98}
R.~Tripp,
Astron. Astrophys. \textbf{331} (1998), 815-820
LBL-40857.

\bibitem [{\citenamefont {Simon}\ \emph {et~al.}(2005)\citenamefont {Simon},
  \citenamefont {Verde},\ and\ \citenamefont {Jimenez}}]{Simon2004}%
  \BibitemOpen
\bibfield  {number} {  }\bibfield  {author} {\bibinfo {author} {\bibfnamefont
  {J.}~\bibnamefont {Simon}}, \bibinfo {author} {\bibfnamefont
  {L.}~\bibnamefont {Verde}},\ and\ \bibinfo {author} {\bibfnamefont
  {R.}~\bibnamefont {Jimenez}},\ }\href@noop {} {\bibfield  {journal} {\bibinfo
   {journal} {Phys. Rev. D}\ }\textbf {\bibinfo {volume} {71}},\ \bibinfo
  {pages} {123001} (\bibinfo {year} {2005})}
\bibitem [{\citenamefont {Stern}\ \emph {et~al.}(2010)\citenamefont {Stern},
  \citenamefont {Jimenez}, \citenamefont {Verde}, \citenamefont
  {Kamionkowski},\ and\ \citenamefont {Stanford}}]{Stern2009}%
  \BibitemOpen
  \bibfield  {author} {\bibinfo {author} {\bibfnamefont {D.}~\bibnamefont
  {Stern}}, \bibinfo {author} {\bibfnamefont {R.}~\bibnamefont {Jimenez}},
  \bibinfo {author} {\bibfnamefont {L.}~\bibnamefont {Verde}}, \bibinfo
  {author} {\bibfnamefont {M.}~\bibnamefont {Kamionkowski}},\ and\ \bibinfo
  {author} {\bibfnamefont {S.~A.}\ \bibnamefont {Stanford}},\ }\href@noop {}
  {\bibfield  {journal} {\bibinfo  {journal} {JCAP}\ }\textbf {\bibinfo
  {volume} {2010}}\bibinfo  {number} { (02)},\ \bibinfo {pages}
  {008}}\BibitemShut {NoStop}%
\bibitem [{\citenamefont {Moresco}\ \emph {et~al.}(2012)\citenamefont {Moresco}
  \emph {et~al.}}]{Moresco2012}%
  \BibitemOpen
\bibfield  {number} {  }\bibfield  {author} {\bibinfo {author} {\bibfnamefont
  {M.}~\bibnamefont {Moresco}} \emph {et~al.},\ }\href@noop {} {\bibfield
  {journal} {\bibinfo  {journal} {JCAP}\ }\textbf {\bibinfo {volume}
  {2012}}\bibinfo  {number} { (08)},\ \bibinfo {pages} {006}}\BibitemShut
  {NoStop}%
\bibitem [{\citenamefont {Zhang}\ \emph {et~al.}(2014)\citenamefont {Zhang},
  \citenamefont {Zhang}, \citenamefont {Yuan}, \citenamefont {Zhang},\ and\
  \citenamefont {Sun}}]{Zhang2012}%
  \BibitemOpen
\bibfield  {number} {  }\bibfield  {author} {\bibinfo {author} {\bibfnamefont
  {C.}~\bibnamefont {Zhang}}, \bibinfo {author} {\bibfnamefont
  {H.}~\bibnamefont {Zhang}}, \bibinfo {author} {\bibfnamefont
  {S.}~\bibnamefont {Yuan}}, \bibinfo {author} {\bibfnamefont {T.-J.}\
  \bibnamefont {Zhang}},\ and\ \bibinfo {author} {\bibfnamefont {Y.-C.}\
  \bibnamefont {Sun}},\ }\href@noop {} {\bibfield  {journal} {\bibinfo
  {journal} {Res. Astron. Astrophys.}\ }\textbf {\bibinfo {volume} {14}},\
  \bibinfo {pages} {1221} (\bibinfo {year} {2014})}\BibitemShut {NoStop}%
\bibitem [{\citenamefont {Moresco}(2015)}]{Moresco2015}%
  \BibitemOpen
  \bibfield  {author} {\bibinfo {author} {\bibfnamefont {M.}~\bibnamefont
  {Moresco}},\ }\href@noop {} {\bibfield  {journal} {\bibinfo  {journal} {Mon.
  Not. Roy. Astron. Soc.}\ }\textbf {\bibinfo {volume} {450}},\ \bibinfo
  {pages} {L16} (\bibinfo {year} {2015})}\BibitemShut {NoStop}%
\bibitem [{\citenamefont {Moresco}\ \emph {et~al.}(2016)\citenamefont
  {Moresco}, \citenamefont {Pozzetti}, \citenamefont {Cimatti}, \citenamefont
  {Jimenez}, \citenamefont {Maraston}, \citenamefont {Verde}, \citenamefont
  {Thomas}, \citenamefont {Citro}, \citenamefont {Tojeiro},\ and\ \citenamefont
  {Wilkinson}}]{Moresco2016}%
  \BibitemOpen
  \bibfield  {author} {\bibinfo {author} {\bibfnamefont {M.}~\bibnamefont
  {Moresco}}, \bibinfo {author} {\bibfnamefont {L.}~\bibnamefont {Pozzetti}},
  \bibinfo {author} {\bibfnamefont {A.}~\bibnamefont {Cimatti}}, \bibinfo
  {author} {\bibfnamefont {R.}~\bibnamefont {Jimenez}}, \bibinfo {author}
  {\bibfnamefont {C.}~\bibnamefont {Maraston}}, \bibinfo {author}
  {\bibfnamefont {L.}~\bibnamefont {Verde}}, \bibinfo {author} {\bibfnamefont
  {D.}~\bibnamefont {Thomas}}, \bibinfo {author} {\bibfnamefont
  {A.}~\bibnamefont {Citro}}, \bibinfo {author} {\bibfnamefont
  {R.}~\bibnamefont {Tojeiro}},\ and\ \bibinfo {author} {\bibfnamefont
  {D.}~\bibnamefont {Wilkinson}},\ }\href@noop {} {\bibfield  {journal}
  {\bibinfo  {journal} {JCAP}\ }\textbf {\bibinfo {volume} {2016}}\bibinfo
  {number} { (05)},\ \bibinfo {pages} {014}}

\bibitem{JimenezEtAl03}
R.~Jimenez, L.~Verde, T.~Treu and D.~Stern,
Astrophys. J. \textbf{593} (2003), 622-629
[arXiv:astro-ph/0302560 [astro-ph]].

\bibitem{GoodWeare} J. Goodman,  and  J. Weare, Comm. App. Math. Comp. Sci., (2010) v.5, 1, 65


\bibitem{ForemanMackey13}
  D.~Foreman-Mackey, D.~W.~Hogg, D.~Lang and J.~Goodman,
  Publ.\ Astron.\ Soc.\ Pac.\  {\bf 125} (2013) 306
  [arXiv:1202.3665 [astro-ph.IM]]
  
  \bibitem [{Note1()}]{Note1}%
  \BibitemOpen
\bibfield  {number} {  }\bibinfo {note} {\url{https://emcee.readthedocs.io/en/stable/}}%




\bibitem{Lewis19}
A.~Lewis,
[arXiv:1910.13970 [astro-ph.IM]].
 

\bibitem{ParkRatra18}
C.~G.~Park and B.~Ratra,
Astrophys. Space Sci. \textbf{364} (2019) no.8, 134
[arXiv:1809.03598 [astro-ph.CO]].
  \bibitem [{Note2()}]{Note2}%
  \BibitemOpen
\bibfield  {number} {  }\bibinfo {note} {See \cite{Seikel:2012uu} and \cite{Jesus:2019nnk} for more details on Gaussian Processes}\BibitemShut {NoStop}%
\bibitem [{\citenamefont {Jesus}\ \emph {et~al.}(2022)\citenamefont {Jesus},
  \citenamefont {Valentim}, \citenamefont {Escobal}, \citenamefont {Pereira},\
  and\ \citenamefont {Benndorf}}]{Urec}%
  \BibitemOpen
  \bibfield  {author} {\bibinfo {author} {\bibfnamefont {J.~F.}\ \bibnamefont
  {Jesus}}, \bibinfo {author} {\bibfnamefont {R.}~\bibnamefont {Valentim}},
  \bibinfo {author} {\bibfnamefont {A.~A.}\ \bibnamefont {Escobal}}, \bibinfo
  {author} {\bibfnamefont {S.~H.}\ \bibnamefont {Pereira}},\ and\ \bibinfo
  {author} {\bibfnamefont {D.}~\bibnamefont {Benndorf}},\ }\href@noop {}
  {\bibfield  {journal} {\bibinfo  {journal} {JCAP}\ }\textbf {\bibinfo
  {volume} {2022}}\bibinfo  {number} { (11)},\ \bibinfo {pages}
  {037}}\BibitemShut {NoStop}%
\bibitem [{Note3()}]{Note3}%
  \BibitemOpen
\bibfield  {number} {  }\bibinfo {note} {GaPP is currently available in the auxiliary repository \url{https://github.com/carlosandrepaes/GaPP}}.
  








\end{thebibliography}

%

\end{document}